%% file: kernel.tex
\documentclass[lean]{Draft}

\input{preamble}

\makeatletter
\newcommand\leanlink{\begingroup\catcode`\#=12\relax\@leanlink}
\newcommand\@leanlink[2]{\endgroup
\href{#1}
{\texttt{\detokenize{#2}}}}

\title[Markov kernels in Mathlib's probability library]{Markov kernels in Mathlib's probability library}
\author[R. Degenne]{
  Rémy Degenne}

\authorinfo[R. Degenne]{Univ. Lille, Inria, CNRS, Centrale Lille, UMR 9189-CRIStAL, F-59000 Lille, France}{remy.degenne@inria.fr}

\VOLUME{NA}
\YEAR{NA}
\NUMBER{NA}
\firstpage{1}
\DOI{NA}
\receiveddate{NA}
\finaldate{NA}
\accepteddate{NA}

\begin{abstract}
The probability folder of Mathlib, Lean's mathematical library, makes a heavy use of Markov kernels.
We present their definition and properties and describe the formalization of the disintegration theorem for Markov kernels.
That theorem is used to define conditional probability distributions of random variables as well as posterior distributions.
We then explain how Markov kernels are used in a more unusual way to get a common definition of independence and conditional independence and, following the same principles, to define sub-Gaussian random variables.
Finally, we also discuss the role of kernels in our formalization of entropy and Kullback-Leibler divergence.
\end{abstract}



\begin{document}

\section{Introduction}\label{sec:intro}

In probability theory, we study objects defined on a probability space, which is a measurable space $\Omega$ (a set equipped with a $\sigma$-algebra of measurable sets), on which a probability measure $P$ is defined.
The probability measure assigns a probability to events (measurable subsets of $\Omega$).
The main objects of study are random variables, which are measurable functions from $\Omega$ to other measurable spaces.
We can ask for the probability of a random variable $X : \Omega \to \mathcal{X}$ to take a value in a measurable set of $\mathcal{X}$.
We will also want for example to determine its distribution, or to prove that it is independent from another random variable, or to find its conditional distribution given a third random variable.
All those random variables are usually defined on the same measurable space $\Omega$, which might not be explicitly specified, and it is understood that $\Omega$ can be extended if needed to introduce more random variables.

A Markov kernel $\kappa : \mathcal{X} \rightsquigarrow \mathcal{Y}$ is a function from a measurable space $\mathcal{X}$ to the set of probability measures on a measurable space $\mathcal{Y}$ satisfying a measurability condition.
It can be seen as describing the effect of a random transformation.
In particular, a Markov kernel $\kappa : \mathcal{X} \rightsquigarrow \mathcal{Y}$ from a probability space maps the probability measure $P$ on $\mathcal{X}$ to a probability measure $\kappa \circ_m P$ on $\mathcal{Y}$ (the precise definition of that measure will be given in Section~\ref{sub:composition}).
One can consider a category with measurable spaces as objects and Markov kernels as morphisms.
In that kernel point of view, there is not a single measurable space on which every random variable is defined, and we focus instead on the transformations between them.

The object of this paper is to present the implementation, properties and use of Markov kernels in the probability folder of \mathlib{} \cite{mathlib}, the mathematical library of the \Lean{} programming language and theorem prover \cite{lean,moura2021lean}.
Markov kernels play a more important role in that library than in most mathematical texts: they provide a unified way to formalize various concepts of probability theory.

\paragraph{Structure of the paper}

We first introduce Markov kernels, detail particular kernels and describe the different ways in which they can be composed in Section~\ref{sec:kernels}.
We then present the formalization of the disintegration theorem for kernels and its use to define conditional distributions (Section~\ref{sec:disintegration}).
Having explained the main properties of kernels, we then present two examples of their use in \mathlib{}, first to define conditional independence and independence (Section~\ref{sec:independence}), then to define sub-Gaussian random variables (Section~\ref{sec:subGaussian}).
We finally discuss an application of kernels to entropy and Kullback-Leibler divergence (Section~\ref{sec:info}).

\paragraph{Codebase}

The definitions and theorems presented in this paper were contributed by the author to the probability folder of \mathlib{}, except for the results of Section~\ref{sec:info} on entropy and divergences that describe work by the author in other repositories.
The development of kernels, their properties and the disintegration theorem consist of around 10000 lines of code in the Probability/Kernel folder (excluding notably the Ionescu-Tulcea subfolder that was not added by the author of this paper).
Results about independence can be found in the Probability/Independence folder (around 4000 lines of code), and the definition of sub-Gaussian random variables is in Probability/Moments (around 2500 lines together with properties of moment generating functions).

This paper contains links to definitions and theorems from the \mathlib{} github repository, like for example \leanlink{https://github.com/leanprover-community/mathlib4/blob/dd465758525e4dcf29027e0ebc5d8fd552be361f/Mathlib/Probability/Kernel/Defs.lean#L55}{ProbabilityTheory.Kernel}.
They all point to a recent commit at the time of writing \footnote{commit dd465758525e4dcf29027e0ebc5d8fd552be361f}.
The name \lstinline|ProbabilityTheory.Kernel| refers to the name of the declaration, not to its path: \lstinline|ProbabilityTheory| is a namespace in which we added a \lstinline|Kernel| definition, and that declaration \lstinline|ProbabilityTheory.Kernel| is placed in the file \texttt{Probability/Kernel/Defs.lean} but could be moved to another file without changing its name.

The results discussed in Section~\ref{sec:info} about entropy were contributed by the author to the PFR project \cite{Anderson_Formalization_of_the_2023}.
This is a collaborative formalization of the article \cite{gowers2025conjecture}, in which the authors prove the polynomial Freiman-Ruzsa conjecture of Katalin Marton.
A few remarks in Section~\ref{sec:info} on Kullback-Leibler and Rényi divergences come from the work of the author and Lorenzo Luccioli on a formalization of information-theoretic divergences in \Lean{} \cite{TestingLowerBounds}.
We are in the process of integrating results about divergences into \mathlib{}: at the time of writing, the definition of Kullback-Leibler divergence is in \mathlib{} but only a few properties have been ported.
Entropy and other divergences have not yet been the object of pull requests.

\paragraph{Notations}
$\mathbb{R}_{+,\infty}$ is the set of nonnegative extended real numbers.
$\Omega, \mathcal{T}, \mathcal{U}, \mathcal{X}, \mathcal{Y}, \mathcal{Z}$ denote measurable spaces.
The set of sets of~$\mathcal{X}$ is denoted by $\mathcal{P}(\mathcal{X})$. 
The set of nonnegative measures on a measurable space $\mathcal{X}$ is written $\mathcal{M}(\mathcal{X})$.
A kernel from $\mathcal{X}$ to $\mathcal{Y}$ is denoted by $\mathcal{X} \rightsquigarrow \mathcal{Y}$.
The expectation of a random variable $X$ with respect to a measure $\mu$ is written $\mu[X] = \int_\omega X(\omega) \, d\mu$.
This is a Lebesgue integral if $X$ is takes values in $\mathbb{R}_{+,\infty}$, and a Bochner integral in other cases (see \cite{gouezel2022formalization} for a discussion of integrals in \mathlib{}).
The notation $f : \mathcal{X} \to \mathcal{Y} \to \mathcal{Z}$ means that $f$ is a curried function $\mathcal{X} \to (\mathcal{Y} \to \mathcal{Z})$ and we write its application to $x$ and $y$ as $f(x)(y)$.
Other notation, notably about constructions of kernels, will be introduced with the corresponding definitions.

\section{Probability spaces and Markov kernels}\label{sec:kernels}

In \mathlib{}, a type can be equipped with a measurable space structure by writing the code \lstinline|{Ω : Type*} [MeasurableSpace Ω]|.
The code snippets in this paper will most often omit the type arguments for brevity: the reader should assume that every type is equipped with a measurable space structure.
A probability measure can be defined on a measurable space with \lstinline|{P : Measure Ω} [IsProbabilityMeasure P]|.

The central object of this paper is the transition kernel (simply kernel in the following) between measurable spaces, and in particular Markov kernels.
A kernel $\kappa : \mathcal{X} \rightsquigarrow \mathcal{Y}$ is a function from $\mathcal{X}$ to the set of measures on $\mathcal{Y}$, such that for every measurable set $B$ of $\mathcal{Y}$, the function $\mathcal{X} \to \mathbb{R}_{+,\infty}$ defined by $x \mapsto \kappa(x)(B)$ is measurable.
It is said to be a Markov kernel if $\kappa(x)$ is a probability measure for every $x \in \mathcal{X}$.

It is useful to think of a Markov kernel as a conditional distribution, in the sense that $\kappa(x)$ is the distribution of some random element of $\mathcal{Y}$ given that a random element of $\mathcal{X}$ takes the value $x$.
Kernels arise naturally when describing the outcomes of sequential experiments.
For a toy example, suppose that if the weather is good one day, we know that it will be good the next day with probability 0.8, and if it is bad weather instead, it will be good the next day with probability 0.4.
We can describe that situation with a Markov kernel $\kappa : \{\text{good}, \text{bad}\} \rightsquigarrow \{\text{good}, \text{bad}\}$ such that $\kappa(\text{good})$ is the probability measure on $\{\text{good}, \text{bad}\}$ giving probability 0.8 to good weather and 0.2 to bad weather, and $\kappa(\text{bad})$ is the probability measure giving probability 0.4 to good weather and 0.6 to bad weather.
On such a simple example there is no need for the measurability condition of kernels: every function from a finite set with discrete $\sigma$-algebra to a measurable space is measurable.
However, the measurability is important in non-discrete spaces.

We can define a measurable space on the set of measures on $\mathcal{Y}$ as the coarsest $\sigma$-algebra such that the function $\mathcal{M}(\mathcal{Y}) \to \mathbb{R}_{+,\infty}$ defined by $\mu \mapsto \mu(B)$ is measurable for every measurable set $B$ of $\mathcal{Y}$.
That measurable space was already implemented in \mathlib{} by Johannes Hölzl in 2019 in the context of the implementation of the Giry monad (see \cite{holzl2017markov} for his work on the same topic in Isabelle/HOL).
Mathematically, a kernel is then simply a measurable function from $\mathcal{X}$ to the set of measures on $\mathcal{Y}$.
In that sense, kernels were already defined by Hötzl in 2019, and the composition of a measure and a kernel (see Section~\ref{sub:composition}) was implemented as the monadic bind of the Giry monad.
But apart from a short file containing the definition and that composition, kernels were not used prior to our work.

We introduced a definition \leanlink{https://github.com/leanprover-community/mathlib4/blob/dd465758525e4dcf29027e0ebc5d8fd552be361f/Mathlib/Probability/Kernel/Defs.lean#L55}{ProbabilityTheory.Kernel} in \mathlib{} as the following structure:

\begin{lstlisting}
structure Kernel ( 𝓧 𝓨 : Type*) [MeasurableSpace 𝓧] [MeasurableSpace 𝓨] where
  toFun : 𝓧 → Measure 𝓨
  measurable' : Measurable toFun
\end{lstlisting}
So \lstinline|Kernel 𝓧 𝓨| is the code for what we write $\mathcal{X} \rightsquigarrow \mathcal{Y}$ in this text.
It is then equipped with a coercion to a function, which means that we can write \lstinline|κ x| for the measure image of $x$ by $\kappa$ instead of \lstinline|κ.toFun x|.
The measurability condition is called \lstinline|measurable'| in the structure to reserve the name \lstinline|measurable| for the more useful measurability of the coercion to a function.

Kernels are fully specified by their action on measurable functions.
That is, if two kernels $\kappa, \eta : \mathcal{X} \rightsquigarrow \mathcal{Y}$ are such that for every measurable function $f : \mathcal{Y} \to \mathbb{R}_{+,\infty}$ and every $x \in \mathcal{X}$, $\int_y f(y) d\kappa(x) = \int_y f(y) d\eta(x)$, then $\kappa = \eta$.
\begin{lstlisting}
theorem ext_fun_iff : κ = η ↔ ∀ x f, Measurable f → ∫⁻ b, f b ∂κ x = ∫⁻ b, f b ∂η x
\end{lstlisting}
The notation \lstinline|∫⁻| indicates that this is a Lebesgue integral.

We will use that fact repeatedly in the following in order to describe kernels by giving their integrals.
Another way to show that two kernels are equal is to show that their values $\kappa(x)$ and $\eta(x)$ are equal for all $x$, which since they are measures can be checked by checking that they give the same value to all measurable sets.

\subsection{Classes of kernels}

A kernel $\kappa : \mathcal{X} \rightsquigarrow \mathcal{Y}$ is a Markov kernel if $\kappa(x)$ is a probability measure for every $x \in \mathcal{X}$.
If the supremum of $\kappa(x)(\mathcal{Y})$ over $x \in \mathcal{X}$ is finite, then $\kappa$ is said to be a finite kernel.
Finally, we also defined the class of s-finite kernels, which are kernels that can be expressed as a countable sum of finite kernels.
Those three properties are denoted by typeclasses \lstinline|[IsMarkovKernel κ]|, \lstinline|[IsFiniteKernel κ]| and \lstinline|[IsSFiniteKernel κ]| respectively.
\cite{vakar2018s} is a good reference on classes of kernels, in particular the less often used s-finite class.

There is a fourth class of kernels that is mentioned in the literature but that we did not implement in \mathlib{}: $\sigma$-finite kernels.
Those are sums of countably many finite kernels that are pairwise singular.
They do not enjoy the nice composition properties of the other classes (see the next subsection), and we decided to wait for a concrete use case before implementing them.
Although, as we will discuss further down, the Radon-Nikodym theorem for kernels stated below is a potential use case.

A measure can be seen as a kernel from the one-point space $\mathbf{1}$ to the space on which it is defined.
A random variable $X : \mathcal{X} \to \mathcal{Y}$ can be seen as a kernel $D_X : \mathcal{X} \rightsquigarrow \mathcal{Y}$ that maps $x$ to the Dirac measure at $X(x)$.
Such a kernel is called a deterministic kernel and is a Markov kernel.
Among the deterministic kernels, a few play a special role.
The identity kernel $\mathrm{id}_{\mathcal{X}} : \mathcal{X} \rightsquigarrow \mathcal{X}$ is the deterministic kernel associated with the identity function: it maps $x$ to the Dirac measure at $x$.
The copy kernel $\Delta_{\mathcal{X}} : \mathcal{X} \rightsquigarrow \mathcal{X} \times \mathcal{X}$ maps $x$ to the Dirac measure at $(x,x)$.
The discard kernel $!_{\mathcal{X}} : \mathcal{X} \rightsquigarrow \mathbf{1}$ maps $x$ to the unique probability measure on the one-point space $\mathbf{1}$.
Another useful (not deterministic) kernel is the constant kernel $\Omega \rightsquigarrow \mathcal{X}$ which maps every point of $\Omega$ to the same probability measure $\mu$ on $\mathcal{X}$.

\subsection{Composition of kernels}\label{sub:composition}

Kernels can be assembled in various ways. The first of those is the composition, which from $\kappa : \mathcal{X} \rightsquigarrow \mathcal{Y}$ and $\eta : \mathcal{Y} \rightsquigarrow \mathcal{Z}$ produces a kernel $\eta \circ_k \kappa : \mathcal{X} \rightsquigarrow \mathcal{Z}$ (\leanlink{https://github.com/leanprover-community/mathlib4/blob/dd465758525e4dcf29027e0ebc5d8fd552be361f/Mathlib/Probability/Kernel/Composition/Comp.lean#L50}{Kernel.comp}).
It is such that for all measurable functions $f : \mathcal Z \to \mathbb{R}_{+,\infty}$ and all $x \in \mathcal X$,
\begin{align*}
\int_z f(z) d(\eta \circ_k \kappa)(x) = \int_y \int_z f(z) d\eta(y) d\kappa(x) \: .
\end{align*}
We use the notation $\circ_k$ to reflect the notation in \mathlib{}, in which we cannot use simply $\circ$ since this is notation for function composition: a kernel $\mathcal{X} \rightsquigarrow \mathcal{Y}$ is a function, but from $\mathcal{X}$ to $\mathcal{M}(\mathcal{Y})$, and composition of kernels is not the same as their composition as functions.

To explain how $\eta \circ_k \kappa$ is defined, we need to first introduce the composition of a kernel and a measure, which we denote by $\kappa \circ_m \mu$ for a kernel $\kappa : \mathcal{X} \rightsquigarrow \mathcal{Y}$ and a measure $\mu$ on $\mathcal{X}$.
This is defined as the bind in the Giry monad \cite{giry1982categorical}, which was implemented prior to our work (\leanlink{https://github.com/leanprover-community/mathlib4/blob/dd465758525e4dcf29027e0ebc5d8fd552be361f/Mathlib/MeasureTheory/Measure/GiryMonad.lean#L225-L228}{Measure.bind}).
The map or pushforward measure of $\mu$ on $\mathcal{X}$ by a measurable function $X : \mathcal{X} \to \mathcal{Z}$, which we denote by $X_* \mu$, is the measure such that $X_*\mu (B) = \mu(X^{-1}(B))$ for all measurable sets $B \subseteq \mathcal{Z}$.
For $\kappa$ and $\mu$, we can thus define the map of $\mu$ by $\kappa$ (seeing $\kappa$ as a measurable function from $\mathcal{X}$ to $\mathcal{M}(\mathcal{Y})$), the measure $\kappa_* \mu$ on $\mathcal{M}(\mathcal{Y})$.
We can then define the join in the monad, which maps a measure $\nu$ on $\mathcal{M}(\mathcal{Y})$ to the measure $\nu^\sharp$ on $\mathcal{Y}$ defined by $\nu^\sharp(B) = \int_\mu \mu(B) d\nu$ for all measurable sets $B$ of $\mathcal{Y}$.
The bind $\kappa \circ_m \mu$ is then defined as the join of the map of $\mu$ by $\kappa$: $\kappa \circ_m \mu = (\kappa_* \mu)^\sharp$.
We can check that map, join and bind are measurable. We don't detail the measurability proof here and refer the reader to \cite{affeldt2025semantics}.
We will also omit measurability proofs for other constructions of kernels.
The composition $\eta \circ_k \kappa$ is then defined as the kernel such that for all $x$, $(\eta \circ_k \kappa)(x) = \eta \circ_m \kappa(x)$.
\newpage
\begin{lstlisting}
def comp (η : Kernel 𝓨 𝓩) (κ : Kernel 𝓧 𝓨) : Kernel 𝓧 𝓩 where
  toFun a := (κ a).bind η
  measurable' := (Measure.measurable_bind' η.measurable).comp κ.measurable
\end{lstlisting}

A composition of Markov kernels is a Markov kernel, and finite and s-finite kernels are also stable by composition.
$\sigma$-finite kernels (which we did not implement) are not stable by composition.
For finite kernels, the existence of a uniform bound $C < \infty$ such that $\kappa(x)(\mathcal{Y}) < C$ for all $x$ is important to have stability: if we require only that $\kappa(x)(\mathcal{Y}) < \infty$ for all $x$, then the composition of two such kernels would be $\sigma$-finite but not necessarily finite.
The stability of Markov, finite and s-finite kernels will also be true for the other constructions presented below.
For $\kappa : \mathcal{X} \rightsquigarrow \mathcal{Y} \times \mathcal{Z}$, we write $\kappa_{\mathrm{fst}} : \mathcal{X} \rightsquigarrow \mathcal{Y}$ and $\kappa_{\mathrm{snd}} : \mathcal{X} \rightsquigarrow \mathcal{Z}$ for the compositions of $\kappa$ with the deterministic kernels associated with the first and second projections respectively.

We can also form the parallel composition of two s-finite kernels $\kappa : \mathcal{X} \rightsquigarrow \mathcal{Y}$ and $\eta : \mathcal{T} \rightsquigarrow \mathcal{U}$, which is a kernel $\kappa \parallel_k \eta : \mathcal{X} \times \mathcal{T} \rightsquigarrow \mathcal{Y} \times \mathcal{U}$ (\leanlink{https://github.com/leanprover-community/mathlib4/blob/dd465758525e4dcf29027e0ebc5d8fd552be361f/Mathlib/Probability/Kernel/Composition/ParallelComp.lean#L46}{Kernel.parallelComp}) defined by
\begin{align*}
(\kappa \parallel_k \eta)(x,t) = \kappa(x) \otimes \eta(t) \: ,
\end{align*}
in which $\kappa(x) \otimes \eta(t)$ is the product measure of $\kappa(x)$ and $\eta(t)$.
Note that here we need the kernels to be s-finite to ensure that the product measure is a measurable function.
In \mathlib{}, the parallel composition is defined for all kernels, but it is set to the default value 0 if either kernel is not s-finite.
This is a standard trick in formalization to avoid having to carry hypotheses in every expression involving that parallel composition (they will instead be present as hypotheses of the lemmas) \cite{defaultValues, defaultValues2}.
All kernel constructions that require s-finiteness use the same trick.

From those two operations, we can form other useful kernels.
The product of two s-finite kernels $\kappa : \mathcal{X} \rightsquigarrow \mathcal{Y}$ and $\eta : \mathcal{X} \rightsquigarrow \mathcal{Z}$ is the kernel $\kappa \times_k \eta : \mathcal{X} \rightsquigarrow \mathcal{Y} \times \mathcal{Z}$, defined to be equal to $ (\kappa \parallel_k \eta) \circ_k \Delta_{\mathcal{X}}$ (\leanlink{https://github.com/leanprover-community/mathlib4/blob/dd465758525e4dcf29027e0ebc5d8fd552be361f/Mathlib/Probability/Kernel/Composition/Prod.lean#L50}{Kernel.prod}).
It is such that for all $x$, $(\kappa \times_k \eta)(x) = \kappa(x) \otimes \eta(x)$.

\begin{table}
    \centering
    \begin{tabular}{c|l}
        $\kappa \circ_k \eta$ & composition of kernels $\kappa$ and $\eta$ \\
        $\kappa \circ_m \mu$ & composition of a kernel $\kappa$ and a measure $\mu$ \\
        $X_* \mu$ & pushforward of a measure $\mu$ by a random variable $X$ \\
        $\kappa \parallel_k \eta$ & parallel composition of kernels $\kappa$ and $\eta$ \\
        $\kappa \otimes_k \eta$ & composition-product of kernels $\kappa$ and $\eta$ \\
        $\mu \otimes_m \kappa$ & composition-product of a measure $\mu$ and a kernel $\kappa$ \\
        $\kappa \times_k \eta$ & product of kernels $\kappa$ and $\eta$
    \end{tabular}
    \caption{Summary of the notations for kernels and measures.}
\end{table}

A more complex construction gives a kernel $\kappa \otimes_k \eta : \mathcal{X} \rightsquigarrow \mathcal{Y} \times \mathcal{Z}$ from two s-finite kernels $\kappa : \mathcal{X} \rightsquigarrow \mathcal{Y}$ and $\eta : \mathcal{X} \times \mathcal{Y} \rightsquigarrow \mathcal{Z}$.
This is called composition-product of kernels in \mathlib{} (\leanlink{https://github.com/leanprover-community/mathlib4/blob/dd465758525e4dcf29027e0ebc5d8fd552be361f/Mathlib/Probability/Kernel/Composition/CompProd.lean#L69}{Kernel.compProd}).
In the literature it is often called simply either composition or product depending on the source, but in \Lean{} code we have to use distinct names.
Intuitively, that operation corresponds to generating data sequentially: first a random element $X$ is observed, then $\kappa$ describes the conditional distribution of $Y$ given $X$, and finally $\eta$ describes the distribution of $Z$ given $(X,Y)$.
We can combine $\kappa$ and $\eta$ into a single kernel that gives the joint distribution of $(Y,Z)$ given $X$. This is $\kappa \otimes_k \eta$.
It is defined as
\begin{align*}
    \kappa \otimes_k \eta
    = D_{s_{\mathcal{Z},\mathcal{Y}}} \circ_k (\eta \parallel_k \id)
    \circ_k D_{\mathrm{assoc}_{\mathcal{X},\mathcal{Y},\mathcal{Y}}}
    \circ_k (\id \parallel_k \Delta_{\mathcal{Y}}) \circ_k (\id \parallel_k \kappa) \circ_k \Delta_{\mathcal{X}}
\end{align*}
in which $s_{\mathcal{Z},\mathcal{Y}} : \mathcal{Z} \times \mathcal{Y} \to \mathcal{Y} \times \mathcal{Z}$ is the function that swaps the two coordinates, and $\mathrm{assoc}_{\mathcal{X},\mathcal{Y},\mathcal{Y}}$ is the associator $\mathcal{X} \times (\mathcal{Y} \times \mathcal{Y}) \to (\mathcal{X} \times \mathcal{Y}) \times \mathcal{Y}$.
More importantly than the formula used in the definition, the composition-product satisfies the property that for all measurable functions $f : \mathcal{Y} \times \mathcal{Z} \to \mathbb{R}_{+,\infty}$ and all $x \in \mathcal{X}$,
\begin{align*}
    \int_{(y,z)} f(y,z) d(\kappa \otimes_k \eta)(x)
    = \int_y \int_z f(y,z) d\eta(x,y) d\kappa(x)
    \: .
\end{align*}

Composition of kernels is associative, $\xi \circ_k (\eta \circ_k \kappa) = (\xi \circ_k \eta) \circ_k \kappa$, and the name ``identity kernel'' is justified by the fact that it is a neutral element for that composition, $\mathrm{id} \circ_k \kappa = \kappa$ and $\kappa \circ_k \mathrm{id} = \kappa$.
The composition-product is, mathematically speaking, associative as well, but here the type theory used in \Lean{} makes it more complex to express that property.
Indeed, suppose that we have three s-finite kernels $\kappa : \mathcal{T} \rightsquigarrow \mathcal{X}$, $\eta : \mathcal{T} \times \mathcal{X} \rightsquigarrow \mathcal{Y}$ and $\xi : \mathcal{T} \times (\mathcal{X} \times \mathcal{Y}) \rightsquigarrow \mathcal{Z}$.
Then we can form a kernel $(\kappa \otimes_k \eta) \otimes_k \xi$ without issue.
Associativity would be the statement that this is the same kernel as $\kappa \otimes_k (\eta \otimes_k \xi)$.
However, there are problems with that second kernel: we cannot build $\eta \otimes_k \xi$ since the type of $\xi$ is $\mathcal{T} \times (\mathcal{X} \times \mathcal{Y}) \rightsquigarrow \mathcal{Z}$ and not $(\mathcal{T} \times \mathcal{X}) \times \mathcal{Y} \rightsquigarrow \mathcal{Z}$.
Those two product of types cannot be used interchangeably.
Similarly, even if we could obtain $\kappa \otimes_k (\eta \otimes_k \xi)$, its type would be $\mathcal{T} \rightsquigarrow \mathcal{X} \times (\mathcal{Y} \times \mathcal{Z})$ while the type of $(\kappa \otimes_k \eta) \otimes_k \xi$ is $\mathcal{T} \rightsquigarrow (\mathcal{X} \times \mathcal{Y}) \times \mathcal{Z}$.
We can insert explicitly compositions with associators to fix the type issues, but that makes the statement of associativity more cumbersome.

Since measures can be seen as kernels from the one-point space, we can use these operations to combine measures and kernels.
From $\mu$ a measure on $\mathcal{X}$ we can build a kernel $c_\mu : \mathbf{1} \rightsquigarrow \mathcal{X}$ that maps the unique point of $\mathbf{1}$ to $\mu$.
We can then compose $c_\mu$ and $\kappa : \mathcal{X} \rightsquigarrow \mathcal{Y}$ to get a kernel $\kappa \circ_k c_\mu : \mathbf{1} \rightsquigarrow \mathcal{Y}$, which takes a single value and defines a measure on $\mathcal{Y}$.
This recovers the composition definition $\kappa \circ_m \mu$ for a kernel $\kappa : \mathcal{X} \rightsquigarrow \mathcal{Y}$ and a measure $\mu$ on $\mathcal{X}$.
By similarly seeing measures as kernels from the one-point space, we can define a composition-product $\mu \otimes_m \kappa$ for a measure $\mu$ on $\mathcal{X}$ and a kernel $\kappa : \mathcal{X} \rightsquigarrow \mathcal{Y}$, which is a measure on $\mathcal{X} \times \mathcal{Y}$ (\leanlink{https://github.com/leanprover-community/mathlib4/blob/dd465758525e4dcf29027e0ebc5d8fd552be361f/Mathlib/Probability/Kernel/Composition/MeasureCompProd.lean#L43}{Measure.compProd}).
For a deterministic kernel $D_X$ associated with a random variable $X : \mathcal{X} \to \mathcal{Y}$, $D_X \circ_m \mu$ is the pushforward measure of $\mu$ by $X$, the measure $X_* \mu$ on $\mathcal{Y}$.

The composition is related to the composition-product by the formula $\eta \circ_k \kappa = (\kappa \otimes_k \eta)_{\mathrm{snd}}$ (in which abuse notation: we see $\eta : \mathcal{Y} \rightsquigarrow \mathcal{Z}$ as a kernel $\mathcal{X} \times \mathcal{Y} \rightsquigarrow \mathcal{Z}$ by ignoring the first component).
In a first attempt at implementing kernels in \mathlib{}, we defined the composition-product as the primitive operation on kernels, and derived the composition and the product from it.
This had the advantage that once the work is done to prove that the composition-product is indeed a kernel, the other operations are easy to define.
There is however a drawback: the composition-product requires s-finite kernels, while the composition should be defined for all kernels.
Defining the composition as $(\kappa \otimes_k \eta)_{\mathrm{snd}}$ is a loss in generality.
The change to use composition and parallel composition as basic operations also aligns better with the categorical viewpoint on kernels.

\subsection{Copy-discard and Markov categories}

We can form several categories with measurable spaces as objects and kernels as morphisms.
Our work on kernels was used by Gaëtan Serré to define copy-discard and Markov categories for kernels in \mathlib{}.
We briefly mention them here to explain how they relate to the objects presented so far.
For an explanation of the different types of categories named below and the theory of Markov categories, see \cite{cho2019disintegration,fritz2020synthetic}.

Measurable spaces and s-finite kernels are a copy-discard category (and in particular a symmetric monoidal category) \leanlink{https://github.com/leanprover-community/mathlib4/blob/dd465758525e4dcf29027e0ebc5d8fd552be361f/Mathlib/Probability/Kernel/Category/SFinKer.lean#L221}{SFinKer} with the following data:
\begin{itemize}
  \item Objects: measurable spaces
  \item Morphisms: s-finite kernels, with identity kernels and composition of kernels $\eta \circ_k \kappa$
  \item Tensor product of objects: product of measurable spaces $\mathcal X \times \mathcal Y$
  \item Tensor product of morphisms: parallel composition $\kappa \parallel_k \eta$ 
  \item Unit: measurable space with one element (denoted by $\mathbf{1}$)
  \item The associator, left and right unitors and braiding are given by the deterministic kernels of the corresponding equivalences between measurable spaces
  \item Copy morphism: $\Delta_{\mathcal X}: \mathcal X \rightsquigarrow \mathcal X \times \mathcal X$
  \item Delete morphism: $!_{\mathcal X} : \mathcal X \rightsquigarrow \mathbf{1}$
\end{itemize}

If we restrict the morphisms to Markov kernels to form the category \leanlink{https://github.com/leanprover-community/mathlib4/blob/dd465758525e4dcf29027e0ebc5d8fd552be361f/Mathlib/Probability/Kernel/Category/Stoch.lean#L59}{Stoch}, we get a Markov category, which is a copy-discard category in which for every morphism $\kappa$, $!_{\mathcal Y} \circ \kappa = !_{\mathcal X}$.

\section{Disintegration and conditional distributions}\label{sec:disintegration}

The previous section showed what kernels are and how they can be composed in different ways.
We now turn to the main theorem about kernels that we formalized in \mathlib{}: the disintegration theorem.

We say that a kernel $\kappa : \mathcal{X} \rightsquigarrow \mathcal{Y} \times \mathcal{Z}$ is disintegrated by a kernel $\eta : \mathcal{X} \times \mathcal{Y} \rightsquigarrow \mathcal{Z}$ if $\kappa = \kappa_{\mathrm{fst}} \otimes_k \eta$.
We also say that $\eta$ is a conditional kernel for $\kappa$.
In the particular case of a product measure $\rho$ on $\mathcal{Y} \times \mathcal{Z}$ (which corresponds to the special case where $\mathcal{X}$ the unit space), disintegration means that $\rho$ can be expressed with the measure $\rho_{\mathrm{fst}}$ on $\mathcal{Y}$ and a kernel $\eta : \mathcal{Y} \rightsquigarrow \mathcal{Z}$ as $\rho = \rho_{\mathrm{fst}} \otimes_m \eta$~: a joint measure can be decomposed into a marginal measure and a conditional distribution.

In general measurable spaces, disintegration is not always possible (refer to \cite{faden1985existence} for exact hypotheses). However, it is possible for every finite kernel if we add assumptions on the spaces.

\begin{definition}[Assumptions on measurable spaces]\label{def:standardBorel}
A measurable space $\mathcal{X}$ has a countably generated $\sigma$-algebra if there exists a countable collection of measurable sets such that the $\sigma$-algebra of $\mathcal{X}$ is the smallest $\sigma$-algebra containing all those sets.

A standard Borel space is a measurable space such that there exists a Polish topology (second-countable topology that can be endowed with a metric for which it is complete) on the underlying set such that the $\sigma$-algebra is the Borel $\sigma$-algebra of that topology (the $\sigma$-algebra generated by the open sets).
\end{definition}

We note that a standard Borel space has a countably generated $\sigma$-algebra.
Furthermore, any standard Borel space embeds measurably into $\mathbb{R}$ (it is isomorphic as a measurable space to a subset of $\mathbb{R}$).
Most spaces used in applied mathematics are standard Borel.

We first introduced the disintegration in \mathlib{} as a definition: under assumptions on the spaces, we built a kernel $\eta$ disintegrating a given kernel $\kappa$.
Since then, Yaël Dillies and Kalle Kytölä introduced a predicate to express that a kernel disintegrates another.

\begin{lstlisting}
class IsCondKernel (κ : Kernel 𝓧 ( 𝓨 × 𝓩)) (κCond : Kernel ( 𝓧 × 𝓨) 𝓩) : Prop where
  protected disintegrate : κ.fst ⊗ₖ κCond = κ
\end{lstlisting}

The command \lstinline|class| makes this structure inferable by the typeclass system of \Lean{}. The \lstinline|protected| modifier means that even if the namespace 
\lstinline|IsCondKernel| is open, the field \lstinline|disintegrate| can only be accessed with the full name \lstinline|IsCondKernel.disintegrate|, which is useful to avoid name clashes with another lemma named \lstinline|disintegrate| in the library.

The advantage of that predicate is that it allows us to work with disintegrations in more general spaces, if we happen to have another mechanism to know that a kernel disintegrates another.
A conditional kernel is almost everywhere equal to a Markov kernel, and almost everywhere unique. Futhermore, if $\eta$ disintegrates $\kappa$ and $\eta$ is $\kappa_{\mathrm{fst}}$-almost everywhere equal to $\eta'$, then $\eta'$ also disintegrates $\kappa$.
That means that if $\kappa$ admits a disintegration, then there is a true Markov kernel that disintegrates $\kappa$.
All those facts are formalized in \mathlib{}.

The main result of this section is the following theorem, which states that there exists a conditional kernel under assumptions on the spaces.

\begin{theorem}[Kernel disintegration]\label{thm:disintegration}
If $\mathcal{Z}$ is a standard Borel space and either $\mathcal{X}$ is countable or $\mathcal{Y}$ has a countably generated $\sigma$-algebra, then every finite kernel $\mathcal{X} \rightsquigarrow \mathcal{Y} \times \mathcal{Z}$ admits a disintegration.
\end{theorem}

It is formalized in \mathlib{} as a definition \leanlink{https://github.com/leanprover-community/mathlib4/blob/dd465758525e4dcf29027e0ebc5d8fd552be361f/Mathlib/Probability/Kernel/Disintegration/StandardBorel.lean#L405}{Kernel.condKernel} that gives a conditional kernel (constructed following the steps outlined in Section~\ref{sub:disintegration_proof}), which we ensure is Markov, together with an instance of the class \lstinline|IsCondKernel|.

\begin{lstlisting}
instance condKernel.instIsCondKernel [CountableOrCountablyGenerated 𝓧 𝓨]
    (κ : Kernel 𝓧 ( 𝓨 × 𝓩)) [IsFiniteKernel κ] :
    κ.IsCondKernel κ.condKernel
\end{lstlisting}
Once this is done, the particular construction of the conditional kernel is not important anymore, as it is uniquely determined up to almost everywhere equality by the disintegration property.

As a corollary of the disintegration theorem for kernels, if $\mathcal{Z}$ is a standard Borel space then every finite measure on $\mathcal{Y} \times \mathcal{Z}$ admits a disintegration.

The proof of Theorem~\ref{thm:disintegration} goes through the construction of densities for finite kernels, and in particular it also gives a formalization of the Radon-Nikodym theorem for kernels.
In order to state that theorem, we need to introduce the definition of a kernel with a density with respect to another kernel.
First, for $\mu$ a measure on $\mathcal{Y}$ and $f : \mathcal{Y} \to \mathbb{R}_{+,\infty}$ a measurable function, we define a measure denoted by $f \cdot \mu$, the measure with density $f$ with respect to $\mu$, by $(f \cdot \mu)(B) = \int_B f(y) d\mu$ for all measurable sets $B$ of $\mathcal{Y}$ (\leanlink{https://github.com/leanprover-community/mathlib4/blob/dd465758525e4dcf29027e0ebc5d8fd552be361f/Mathlib/MeasureTheory/Measure/WithDensity.lean#L37-L42}{Measure.withDensity}).
For $f : \mathcal{X} \to \mathcal{Y} \to \mathbb{R}_{+,\infty}$ a jointly measurable function (that is, a function such that the uncurried version is measurable on the product $\mathcal{X} \times \mathcal{Y}$) and $\eta : \mathcal{X} \rightsquigarrow \mathcal{Y}$ an s-finite kernel, we define the kernel $f \cdot \eta : \mathcal{X} \rightsquigarrow \mathcal{Y}$ (\leanlink{https://github.com/leanprover-community/mathlib4/blob/dd465758525e4dcf29027e0ebc5d8fd552be361f/Mathlib/Probability/Kernel/WithDensity.lean#L48}{Kernel.withDensity}) such that
\begin{align*}
    (f \cdot \eta)(x)(B) = \int_{y \in B} f(x)(y) d\eta(x)
    \: .
\end{align*}
That is, for all $x$, $(f \cdot \eta)(x)$ is defined as the measure $f(x) \cdot \eta(x)$ on $\mathcal{Y}$, and we can prove that thanks to the joint measurability of $f$ and the s-finiteness of $\eta$, $f \cdot \eta$ is indeed a kernel.
In \mathlib{} we extend the definitions to functions that may not be measurable, by using the default value 0 is $f$ is not jointly measurable.

Recall that a measure $\mu$ is absolutely continuous with respect to a measure $\nu$ if for all measurable sets $B$, $\nu(B) = 0$ implies $\mu(B) = 0$, and is singular with respect to $\nu$ if there exists a measurable set $A$ such that $\mu(A) = 0$ and $\nu(\mathcal{Y} \setminus A) = 0$.

With those definitions, we can state the Radon-Nikodym theorem for kernels.

\begin{theorem}[Radon-Nikodym]\label{thm:radon-nikodym}
Suppose that either $\mathcal{X}$ is countable or $\mathcal{Y}$ has a countably generated $\sigma$-algebra.
Let $\kappa, \eta : \mathcal{X} \rightsquigarrow \mathcal{Y}$ be finite kernels.
Then there exists a jointly measurable function $\frac{d \kappa}{d \eta} : \mathcal{X} \to \mathcal{Y} \to \mathbb{R}_{+,\infty}$ and a finite kernel $\kappa_{\perp \eta} : \mathcal{X} \rightsquigarrow \mathcal{Y}$ such that for every $x \in \mathcal{X}$, $\kappa_{\perp \eta}(x)$ is singular with respect to $\eta(x)$ and $\kappa = \frac{d \kappa}{d \eta} \cdot \eta + \kappa_{\perp \eta}$~.
\end{theorem}

The remarkable feature of Theorem~\ref{thm:radon-nikodym} is the joint measurability of the density $\frac{d \kappa}{d \eta}$ (\leanlink{https://github.com/leanprover-community/mathlib4/blob/dd465758525e4dcf29027e0ebc5d8fd552be361f/Mathlib/Probability/Kernel/RadonNikodym.lean#L228}{Kernel.rnDeriv}).
If we only wanted measurability of $\frac{d \kappa}{d \eta}(x)$ for every $x \in \mathcal{X}$, the proof would be much simpler and would not require the assumption that $\mathcal{Y}$ has a countably generated $\sigma$-algebra: we would simply use the Radon-Nikodym theorem for measures on $\mathcal{Y}$ for each $x \in \mathcal{X}$ separately.
The kernel Radon-Nikodym derivative is almost everywhere equal to the measure Radon-Nikodym derivative $\frac{d \kappa(x)}{d \eta(x)}$ and the main point of the theorem is that we can choose those derivatives (which are defined only almost everywhere) in a jointly measurable way.

Theorem~\ref{thm:radon-nikodym} is stated for finite kernels, but it could be extended to $\kappa$ an s-finite kernel and $\eta$ a $\sigma$-finite kernel (which would require an implementation of $\sigma$-finite kernels).

\subsection{Proof of the disintegration theorem}\label{sub:disintegration_proof}

The main result from which both the disintegration theorem and the Radon-Nikodym theorem are derived is the existence of densities for finite kernels.
Note that there is an unfortunate clash of terminology here: what we call density in this section is not a function such that $\kappa = d \cdot \eta$ for some kernel $\eta$ as in \lstinline|Kernel.withDensity|. It has similar properties (integrating the density with respect to a kernel gives another kernel), but its type is different.
For two kernels $\kappa, \eta : \mathcal{X} \rightsquigarrow \mathcal{Y}$, we write $\kappa \le \eta$ if for all $x \in \mathcal{X}$ and measurable set $B$ of $\mathcal{Y}$, $\kappa(x)(B) \le \eta(x)(B)$.

\begin{theorem}[Existence of densities]\label{thm:density}
Let $\kappa : \mathcal{X} \rightsquigarrow \mathcal{Y} \times \mathcal{Z}$ and $\nu : \mathcal{X} \rightsquigarrow \mathcal{Y}$ be two finite kernels with $\kappa_{\mathrm{fst}} \le \nu$, where $\mathcal{Y}$ has a countably generated $\sigma$-algebra.
Then there exists a function $d_{\kappa, \nu} : \mathcal{X} \to \mathcal{Y} \to \mathcal{P}(\mathcal{Z}) \to \mathbb{R}$ jointly measurable on $\mathcal{X} \times \mathcal{Y}$ such that for all $x \in \mathcal{X}$ and all measurable sets $B$ of $\mathcal{Y}$ and $C$ of $\mathcal{Z}$, $\int_{y \in B} d_{\kappa, \nu}(x)(y)(C) \, d\nu(x) = \kappa(x)(B \times C)$.
\end{theorem}

We added a definition for the density, \leanlink{https://github.com/leanprover-community/mathlib4/blob/dd465758525e4dcf29027e0ebc5d8fd552be361f/Mathlib/Probability/Kernel/Disintegration/Density.lean#L427}{Kernel.density}, and its main property is stated in the following lemma.

\begin{lstlisting}
lemma setLIntegral_density [CountablyGenerated 𝓨] (hκν : fst κ ≤ ν) [IsFiniteKernel ν]
    (a : 𝓧) (hB : MeasurableSet B) (hC : MeasurableSet C) :
    ∫⁻ x in B, ENNReal.ofReal (density κ ν a x C) ∂(ν a) = κ a (B ×ˢ C)
\end{lstlisting}

\paragraph{From densities to disintegration}

Using the fact that a standard Borel space embeds measurably into $\mathbb{R}$ (\leanlink{https://github.com/leanprover-community/mathlib4/blob/dd465758525e4dcf29027e0ebc5d8fd552be361f/Mathlib/MeasureTheory/Constructions/Polish/EmbeddingReal.lean#L63}{embeddingReal}), we can reduce the proof of Theorem~\ref{thm:disintegration} to the case where $\mathcal{Z} = \mathbb{R}$: for a general standard Borel space, we map the kernel to one with values in $\mathbb{R}$, disintegrate it, and then pull back the conditional kernel to the original space.
Thus from a finite kernel $\kappa : \mathcal{X} \rightsquigarrow \mathcal{Y} \times \mathbb{R}$, we want to build a kernel $\eta : \mathcal{X} \times \mathcal{Y} \rightsquigarrow \mathbb{R}$ such that $\kappa = \kappa_{\mathrm{fst}} \otimes_k \eta$.
Since $\eta$ takes values in the measures on $\mathbb{R}$, we can build them from their cumulative distribution functions.
A cumulative distribution function (CDF) is a nondecreasing right-continuous function $F : \mathbb{R} \to \mathbb{R}$ with limit 0 at $-\infty$ and limit 1 at $+\infty$, and characterizes a probability measure on $\mathbb{R}$.
Nondecreasing right-continuous functions $\mathbb{R} \to \mathbb{R}$ are bundled in \mathlib{} in a type \lstinline|StieltjesFunction| and the corresponding measure can be obtained with \leanlink{https://github.com/leanprover-community/mathlib4/blob/dd465758525e4dcf29027e0ebc5d8fd552be361f/Mathlib/MeasureTheory/Measure/Stieltjes.lean#L528}{StieltjesFunction.measure}.

More precisely, here we want to define a conditional kernel CDF for $\kappa : \mathcal{X} \rightsquigarrow \mathcal{Y} \times \mathbb{R}$ with respect to another kernel $\nu : \mathcal{X} \rightsquigarrow \mathcal{Y}$ (and we will apply it to $\nu = \kappa_{\mathrm{fst}}$).
We call a function $f : \mathcal{X} \times \mathcal{Y} \to \mathbb{R} \to \mathbb{R}$ which is nondecreasing right-continuous for all $p \in \mathcal{X} \times \mathcal{Y}$ a conditional kernel CDF of $\kappa : \mathcal{X} \rightsquigarrow \mathcal{Y} \times \mathbb{R}$ with respect to $\nu : \mathcal{X} \rightsquigarrow \mathcal{Y}$ if for all $z \in \mathbb{R}$, $p \mapsto f(p)(z)$ is measurable on the product $\mathcal{X} \times \mathcal{Y}$, if $f(p)$ tends to 0 at $-\infty$ and to 1 at $+ \infty$ for all $p \in \mathcal{X} \times \mathcal{Y}$, if $y \mapsto f (x, y)(z)$ is $\nu(x)$-integrable for all $x \in \mathcal{X}$ and $z \in \mathbb{R}$ and for all measurable sets $B$ of~$\mathcal{Y}$, $\int_{y \in B} f (x, y)(z) \, d\nu(x) = \kappa(x)(B \times (-\infty,z])$.
For each $(x,y) \in \mathcal{X} \times \mathcal{Y}$, $f(x,y)$ is a CDF and thus defines a probability measure on $\mathbb{R}$.
The joint measurability of $f$ ensures that the collection of measures defines a kernel.

Thanks to the continuity properties of CDFs, we can build first a function $f : \mathcal{X} \times \mathcal{Y} \to \mathbb{Q} \to \mathbb{R}$ (note the rational numbers $\mathbb{Q}$ instead of the reals $\mathbb{R}$) satisfying all those properties, and then extend it to a function $\mathcal{X} \times \mathcal{Y} \to \mathbb{R} \to \mathbb{R}$.
The countability of $\mathbb{Q}$ will be helpful when it comes to proving the right-continuity of $f$.

If $\mathcal{X}$ is countable, we can provide for each $x \in \mathcal{X}$ a measurable function $f : \mathcal{Y} \to \mathbb{Q} \to \mathbb{R}$ and proceed as above to obtain a kernel $\mathcal{Y} \rightsquigarrow \mathbb{R}$.
Since $\mathcal{X}$ is countable, measurability is not an issue and we can put those kernels together into a kernel $\mathcal{X} \times \mathcal{Y} \rightsquigarrow \mathbb{R}$.
If on the other hand $\mathcal{X}$ is not countable, we cannot proceed separately for each $x \in \mathcal{X}$ and have to build a function which is measurable on the product.
We are able to do so if $\mathcal{Y}$ has a countably generated $\sigma$-algebra thanks to the density provided by Theorem~\ref{thm:density}.
The density $d_{\kappa, \nu} : \mathcal{X} \to \mathcal{Y} \to \mathcal{P}(\mathbb{R}) \to \mathbb{R}$ is jointly measurable on $\mathcal{X} \times \mathcal{Y}$ and such that for all $x \in \mathcal{X}$ and all measurable sets $B$ of $\mathcal{Y}$ and all $q \in \mathbb{Q}$, $\int_{y \in B} d_{\kappa, \nu}(x)(y)((-\infty,q]) \, d\nu(x) = \kappa(x)(B \times (-\infty,q])$.
It thus satisfies the integral condition, and we are able to check that the other conditions are also fulfilled.

The countability of $\mathbb{Q}$ is important when proving the right-continuity.
We want to prove that for almost all $p \in \mathcal{X} \times \mathcal{Y}$, $d_{\kappa, \nu}(p)$ is right-continuous on $\mathbb{Q}$~.
Since $\mathbb{Q}$ is countable, it is enough to prove that for all $q\in \mathbb{Q}$, for almost all $p$, $d_{\kappa, \nu}(p)$ is right-continuous at $q$.
To prove the almost everywhere equality of the limit on the right and the value of the function at a point $q \in \mathbb{Q}$, it suffices to prove that the integrals of those functions are equal on any measurable set $A$ of $\mathcal{X} \times \mathcal{Y}$.
And since we know the integrals of $d_{\kappa, \nu}$, we can prove the equality of integrals.

\paragraph{From densities to Radon-Nikodym}

From $\kappa, \eta : \mathcal{X} \rightsquigarrow \mathcal{Y}$ two finite kernels we want to build a jointly measurable function $\frac{d \kappa}{d \eta} : \mathcal{X} \to \mathcal{Y} \to \mathbb{R}_{+,\infty}$ and a finite kernel $\kappa_{\perp \eta} : \mathcal{X} \rightsquigarrow \mathcal{Y}$ such that for every $x \in \mathcal{X}$, $\kappa_{\perp \eta}(x)$ is singular with respect to $\eta(x)$ and $\kappa = \frac{d \kappa}{d \eta} \cdot \eta + \kappa_{\perp \eta}$~.

We use Theorem~\ref{thm:density} for $\mathcal{Z} = \mathbf{1}$, the one-point space, to obtain the following corollary.

\begin{corollary}\label{cor:density_rn}
Let $\kappa, \eta : \mathcal{X} \rightsquigarrow \mathcal{Y}$ be two finite kernels with $\kappa \le \eta$, in which $\mathcal{Y}$ has a countably generated $\sigma$-algebra.
Then there exists a function $d_{\kappa, \eta} : \mathcal{X} \to \mathcal{Y} \to \mathbb{R}$ jointly measurable on $\mathcal{X} \times \mathcal{Y}$ such that for all $x \in \mathcal{X}$ and all measurable sets $B$ of $\mathcal{Y}$, $\int_{y \in B} d_{\kappa, \eta}(x)(y) \, d\eta(x) = \kappa(x)(B)$.
\end{corollary}

We use the corollary to define the Radon-Nikodym derivative of $\kappa$ with respect to $\eta$ in two steps.
First, we define an auxiliary function \lstinline|rnDerivAux| which is defined as the density $d_{\kappa, \eta}$ of the Corollary when $\mathcal{Y}$ has a countably generated $\sigma$-algebra and as the element-wise Radon-Nikodym derivative when $\mathcal{X}$ is countable.
\newpage
\begin{lstlisting}
def rnDerivAux [hCoC : CountableOrCountablyGenerated 𝓧 𝓨]
    (κ η : Kernel 𝓧 𝓨) (x : 𝓧) (y : 𝓨) : ℝ :=
  if hC : Countable 𝓧 then ((κ x).rnDeriv (η x) y).toReal
  else haveI := hCoC.countableOrCountablyGenerated.resolve_left hC
    density (map κ (fun a ↦ (a, ()))) η x y univ
\end{lstlisting}
In this auxiliary definition, \lstinline|density| is the function from Theorem~\ref{thm:density} and \lstinline|density (map κ (fun a ↦ (a, ()))) η x y univ| is the function from Corollary~\ref{cor:density_rn} (which in the code is defined even if we don't have $\kappa \le \eta$ but has meaningful properties only in that case).

In the Radon-Nikodym Theorem~\ref{thm:radon-nikodym}, we do not have the condition $\kappa \le \eta$.
We thus apply \lstinline|rnDerivAux| to $\kappa$ and $\kappa + \eta$, which satisfy that condition.
The Radon-Nikodym derivative $\frac{d \kappa}{d \eta}$ (\leanlink{https://github.com/leanprover-community/mathlib4/blob/dd465758525e4dcf29027e0ebc5d8fd552be361f/Mathlib/Probability/Kernel/RadonNikodym.lean#L226-L229}{Kernel.rnDeriv}) is defined by the ratio $\frac{d_{\kappa, \kappa + \eta}}{1 - d_{\kappa, \kappa + \eta}}$.
\begin{lstlisting}
def rnDeriv [CountableOrCountablyGenerated 𝓧 𝓨]
    (κ η : Kernel 𝓧 𝓨) (x : 𝓧) (y : 𝓨) : ℝ≥0∞ :=
  ENNReal.ofReal (rnDerivAux κ (κ + η) x y) / ENNReal.ofReal (1 - rnDerivAux κ (κ + η) x y)
\end{lstlisting}
We also use \lstinline|rnDerivAux| to define the singular part (\leanlink{https://github.com/leanprover-community/mathlib4/blob/dd465758525e4dcf29027e0ebc5d8fd552be361f/Mathlib/Probability/Kernel/RadonNikodym.lean#L257-L262}{Kernel.rnSingularPart}) $\kappa_{\perp \eta}$ of $\kappa$ with respect to $\eta$, as $\kappa_{\perp \eta} = (d_{\kappa, \kappa + \eta} - (1 - d_{\kappa, \kappa + \eta})\frac{d \kappa}{d \eta}) \cdot (\kappa + \eta)$.
With the conventions of Lean regarding division by zero, the term $(1 - d_{\kappa, \kappa + \eta}(x,y))\frac{d \kappa}{d \eta}(x,y) = (1 - d_{\kappa, \kappa + \eta}(x,y)) \frac{d_{\kappa, \kappa + \eta}(x, y)}{1 - d_{\kappa, \kappa + \eta}(x,y)}$ is equal to 0 when $d_{\kappa, \kappa + \eta}(x,y) = 1$ and to $d_{\kappa, \kappa + \eta}(x,y)$ otherwise.
\begin{lstlisting}
def singularPart [CountableOrCountablyGenerated 𝓧 𝓨]
    (κ η : Kernel 𝓧 𝓨) [IsSFiniteKernel κ] [IsSFiniteKernel η] :
    Kernel 𝓧 𝓨 :=
  withDensity (κ + η) (fun x y ↦ Real.toNNReal (rnDerivAux κ (κ + η) x y)
    - Real.toNNReal (1 - rnDerivAux κ (κ + η) x y) * rnDeriv κ η x y)
\end{lstlisting}
It is then not hard to check that $\frac{d \kappa}{d \eta}$ and $\kappa_{\perp \eta}$ satisfy the required properties when the kernels are finite.
Thanks to the Radon-Nikodym Theorem~\ref{thm:radon-nikodym} which is now proved, the condition $\kappa \le \eta$ in Corollary~\ref{cor:density_rn} can be relaxed to absolute continuity of $\kappa$ with respect to $\eta$ (that is, absolute continuity of $\kappa(x)$ with respect to $\eta(x)$ for all $x$).

\paragraph{Proof of the existence of densities}

We now give a sketch of the proof of Theorem~\ref{thm:density}.
Our proof follows \cite[Theorem 9.27]{kallenberg2021}.
Recall that $\kappa : \mathcal{X} \rightsquigarrow \mathcal{Y} \times \mathcal{Z}$ and $\nu : \mathcal{X} \rightsquigarrow \mathcal{Y}$ are two finite kernels with $\kappa_{\mathrm{fst}} \le \nu$, where $\mathcal{Y}$ has a countably generated $\sigma$-algebra.
If we were interested in a single $x \in \mathcal{X}$, we could use the Radon-Nikodym theorem for measures to obtain a density.
For $x \in \mathcal{X}$, $y \in \mathcal{Y}$ and $C$ a measurable set of $\mathcal{Z}$, let $\kappa(x)_{| \mathcal{Y} \times C}$ be the restriction of the measure $\kappa(x)$ to the set $\mathcal{Y} \times C$. Let then $(\kappa(x)_{| \mathcal{Y} \times C})_{\mathrm{fst}}$ be the marginal of that measure on $\mathcal{Y}$.
Then we could define the density as $d'_{\kappa, \nu}(x, y)(C) = \frac{d (\kappa(x)_{| \mathcal{Y} \times C})_{\mathrm{fst}}}{d \nu(x)}(y)$ and it would satisfy the integral condition and be measurable in $y$.
The problem is to obtain a density that is jointly measurable on $\mathcal{X} \times \mathcal{Y}$.

If the $\sigma$-algebra on $\mathcal{Y}$ is finite, then we can proceed as follows.
Let $\{B_1, \ldots, B_k\}$ be a partition of $\mathcal{Y}$ that generates its $\sigma$-algebra and for $y \in \mathcal{Y}$, let $B(y)$ be the set in the partition containing $y$ (this is a measurable function).
We can then define $d_{\kappa, \nu}(x, y)(C) = \frac{\kappa(x)(B(y) \times C)}{\nu(x)(B(y))}$, which is jointly measurable on $\mathcal{X} \times \mathcal{Y}$ and satisfies the integral condition.

Our hypothesis on $\mathcal{Y}$ is not that its $\sigma$-algebra is finite, but that it is countably generated.
But that assumption will allow us to reduce to the finite case.
For a countably generated $\sigma$-algebra, there exists a sequence of finer and finer $\sigma$-algebras $\mathcal{F}_n$, each generated by a partition $\{B_{n,1}, \ldots, B_{n,k_n}\}$, whose union generates the $\sigma$-algebra of $\mathcal{Y}$ (see \leanlink{https://github.com/leanprover-community/mathlib4/blob/dd465758525e4dcf29027e0ebc5d8fd552be361f/Mathlib/MeasureTheory/MeasurableSpace/CountablyGenerated.lean#L449-L453}{countablePartition} for the partition into finitely many sets, and \leanlink{https://github.com/leanprover-community/mathlib4/blob/dd465758525e4dcf29027e0ebc5d8fd552be361f/Mathlib/Probability/Process/PartitionFiltration.lean#L106-L112}{countableFiltration} for the filtration built out of them).

For each $n$, we can define a density $d_{\kappa, \nu}^n$ as above for the finite $\sigma$-algebra $\mathcal{F}_n$, which is jointly measurable on $\mathcal{X} \times \mathcal{Y}$ and satisfies the integral condition for all sets in $\mathcal{F}_n$.

\begin{lstlisting}
def densityProcess (κ : Kernel 𝓧 ( 𝓨 × 𝓩)) (ν : Kernel 𝓧 𝓨)
    (n : ℕ) (x : 𝓧) (y : 𝓨) (C : Set 𝓩) : ℝ :=
  (κ x (countablePartitionSet n y ×ˢ C) / ν x (countablePartitionSet n y)).toReal
\end{lstlisting}
Here \lstinline|countablePartitionSet n y| is the set in the partition $\{B_{n,1}, \ldots, B_{n,k_n}\}$ containing $y$.
The key to turn that sequence into a density is to use a martingale convergence theorem, which was formalized and added to \mathlib{} by \citet{ying2023formalization}.
$(\mathcal{F}_n)_{n \in \mathbb{N}}$ is a filtration, and for each $x \in \mathcal{X}$ and each measurable set $C$ of $\mathcal{Z}$, the sequence of functions $(d_{\kappa, \nu}^n(x, \cdot)(C))_{n \in \mathbb{N}} : \mathbb{N} \to \mathcal{Y} \to \mathbb{R}$ can be shown to be a martingale which is bounded in $L^1$.

\begin{lstlisting}
lemma martingale_densityProcess (hκν : fst κ ≤ ν) [IsFiniteKernel ν]
    (x : 𝓧) {C : Set 𝓩} (hC : MeasurableSet C) :
    Martingale (fun n y ↦ densityProcess κ ν n x y C) (countableFiltration 𝓨) (ν x)
\end{lstlisting}

Therefore, by the martingale convergence theorem, it converges almost everywhere to a limit (\leanlink{https://github.com/leanprover-community/mathlib4/blob/dd465758525e4dcf29027e0ebc5d8fd552be361f/Mathlib/Probability/Martingale/Convergence.lean#L309-L320}{Submartingale.tendsto_eLpNorm_one_limitProcess}).
That limit is almost everywhere equal to the limit superior of the sequence, which is a jointly measurable function.
We define $d_{\kappa, \nu}(x, y)(C)$ to be that limit superior.
\begin{lstlisting}
def density (κ : Kernel 𝓧 ( 𝓨 × 𝓩)) (ν : Kernel 𝓧 𝓨) (x : 𝓧) (y : 𝓨) (C : Set 𝓩) : ℝ :=
  limsup (fun n ↦ densityProcess κ ν n x y C) atTop
\end{lstlisting}
It is jointly measurable on $\mathcal{X} \times \mathcal{Y}$ and since it is almost everywhere the limit of the density process in $L^1$, it satisfies the integral condition for all measurable sets $C$ of $\mathcal{Z}$.

\subsection{Conditional distributions}\label{sub:condDistrib}

The application of disintegration to random variables gives the notion of conditional distribution.
Let $(\Omega, P)$ be a probability space, and let $X : \Omega \to \mathcal{X}$ and $Y : \Omega \to \mathcal{Y}$ be two random variables.
We can consider the joint distribution of $(X,Y)$, which is the probability measure $(X, Y)_*P$ on $\mathcal{X} \times \mathcal{Y}$ (pushforward of $P$ by the map $(X, Y)$) that we denote by $P_{XY}$.
The distribution of $X$ is likewise denoted by $P_X$.
If $\mathcal{Y}$ is a standard Borel space, then by the disintegration theorem there exists a kernel $P_{Y|X} : \mathcal{X} \rightsquigarrow \mathcal{Y}$ such that $P_{XY} = P_X \otimes_m P_{Y|X}$.
That kernel is called a (regular) conditional probability distribution of $Y$ given $X$.
This disintegration is exactly how we define conditional distributions in Mathlib (\leanlink{https://github.com/leanprover-community/mathlib4/blob/dd465758525e4dcf29027e0ebc5d8fd552be361f/Mathlib/Probability/Kernel/CondDistrib.lean#L54-L62}{condDistrib}).

\begin{lstlisting}
def condDistrib (Y : Ω → 𝓨) (X : Ω → 𝓧) (μ : Measure Ω) [IsFiniteMeasure μ] :
    Kernel 𝓧 𝓨 :=
  (μ.map fun a ↦ (X a, Y a)).condKernel
\end{lstlisting}

As an important special case of a conditional distribution, in a standard Borel space we can define a Markov kernel associated to a conditional expectation.
Let $\mathcal{G}$ be a sub-$\sigma$-algebra of the $\sigma$-algebra $\mathcal{F}$ of $\Omega$.
Let $E$ be a Banach space.
The conditional expectation of an integrable function $f : \Omega \to E$ with respect to $\mathcal{G}$ is a $\mathcal{G}$-measurable function $P[f | \mathcal{G}] : \Omega \to E$ such that for every $\mathcal{G}$-measurable set $A$, $\int_A P[f | \mathcal{G}] dP = \int_A f dP$.
It exists and is almost everywhere unique.
For a measurable set $B$, we write $P(B | \mathcal{G})$ for $P[\mathbf{1}_B | \mathcal{G}]$.

In a standard Borel space, we can obtain the conditional distributions for the functions $Y = \mathrm{id}_{\Omega}$ (from $(\Omega, \mathcal{F})$ to itself) and $X$ the identity function on $\Omega$ but seen as a function from $(\Omega, \mathcal{F})$ to $(\Omega, \mathcal{G})$.
This is a kernel from $(\Omega, \mathcal{G})$ to $(\Omega, \mathcal{F})$, which we denote by $P^{\mathcal{G}}$ and is called \leanlink{https://github.com/leanprover-community/mathlib4/blob/dd465758525e4dcf29027e0ebc5d8fd552be361f/Mathlib/Probability/Kernel/Condexp.lean#L71}{condExpKernel} in \mathlib{}.
Its main property is that for every integrable function $f : \Omega \to \mathbb{R}$, the conditional expectation $P[f | \mathcal{G}]$ is almost everywhere equal to the function $\omega \mapsto P^{\mathcal{G}}(\omega)[f]$.

\begin{lstlisting}
theorem condExp_ae_eq_integral_condExpKernel {m mΩ : MeasurableSpace Ω}
    {μ : Measure Ω} [IsFiniteMeasure μ] [NormedAddCommGroup E] {f : Ω → E}
    [NormedSpace ℝ E] [CompleteSpace E] (hm : m ≤ mΩ) (hf_int : Integrable f μ) :
    μ[f|m] =ᵐ[μ] fun ω => ∫ y, f y ∂condExpKernel μ m ω
\end{lstlisting}

That code uses the notation \lstinline|f =ᵐ[μ] g|, which means that $f$ and $g$ are equal $\mu$-almost everywhere (the character \lstinline|ᵐ| is part of the notation and is not related to the measurable space \lstinline|m|).

In general the conditional expectations $P(A | \mathcal{G})$ and $P(B | \mathcal{G})$ for different sets $A$ and $B$ are two functions defined only almost everywhere, and for example we have only almost everywhere the equality $P(A \cup B | \mathcal{G}) = P(A | \mathcal{G}) + P(B | \mathcal{G})$ for $A,B$ disjoint.
We can go around that issue as long as only countably many sets are involved, but having a unique version of the conditional expectation that consolidates into a kernel (\lstinline|condExpKernel|) is much more convenient since it allows the use of equality instead of almost-everywhere equality.

\subsection{Posterior kernel}

A measure $\mu$ on $\mathcal{X}$ and a kernel $\kappa : \mathcal{X} \rightsquigarrow \mathcal{Y}$ define a joint measure $\mu \otimes_m \kappa$ on $\mathcal{X} \times \mathcal{Y}$.
We already know a disintegration of that measure by construction: it is disintegrated by $\kappa$.
However we can map the measure by the function that swaps the two coordinates, and we obtain a joint measure on $\mathcal{Y} \times \mathcal{X}$.
The marginal of that measure on $\mathcal{Y}$ is $\kappa \circ_m \mu$.
If $\mathcal{X}$ is a standard Borel space, then by the disintegration theorem there exists a kernel $\kappa_{\mu}^\dagger : \mathcal{Y} \rightsquigarrow \mathcal{X}$ such that the swapped joint measure $s_{\mathcal{X} \times \mathcal{Y} *}(\mu \otimes_m \kappa)$ is equal to $(\kappa \circ_m \mu) \otimes_m \kappa_{\mu}^\dagger$.

The kernel $\kappa_{\mu}^\dagger$ is called the posterior kernel or Bayesian inverse of $\kappa$ with respect to $\mu$.
It is a central object in Bayesian statistics, where $\mu$ is interpreted as the prior distribution on a parameter, $\kappa$ gives the distribution of the observed data given the parameter, and $\kappa_{\mu}^\dagger$ maps data to a distribution on the parameter called posterior distribution, that is interpreted as the updated belief about the parameter after observing the data.

A remark about notation: the notation $\kappa_{\mu}^\dagger$ makes the measure explicit, which departs from the notation $\kappa^\dagger$ used in parts of the literature on posterior kernels \cite{clerc2017pointless}, in which the measure is understood from the context.
In the code, we have to keep the measure explicit because it cannot be inferred otherwise, and we chose to keep the notation of this paper close to the code.

\begin{lstlisting}
def posterior (κ : Kernel Ω 𝓧) (μ : Measure Ω) [IsFiniteMeasure μ] [IsFiniteKernel κ] :
    Kernel 𝓧 Ω :=
  ((μ ⊗ₘ κ).map Prod.swap).condKernel

scoped[ProbabilityTheory] infix:arg "†" => ProbabilityTheory.posterior
\end{lstlisting}

The last line defines a notation (available only in the \lstinline|ProbabilityTheory| namespace) that makes it possible to write $\kappa_{\mu}^\dagger$ in code as \lstinline|κ†μ|.
As a check that the definition is correct, we can prove that the posterior kernel satisfies Bayes' theorem.

\begin{lstlisting}
lemma posterior_eq_withDensity (h_ac : ∀ᵐ ω ∂μ, κ ω ≪ κ ∘ₘ μ) :
    ∀ᵐ y ∂(κ ∘ₘ μ), (κ†μ) y
      = μ.withDensity (fun x ↦ κ.rnDeriv (Kernel.const _ (κ ∘ₘ μ)) x y)
\end{lstlisting}

That lemma states that for $\kappa \circ_m \mu$-almost all $y \in \mathcal{Y}$ (written \lstinline|∀ᵐ y ∂(κ ∘ₘ μ)| in code), $\kappa_{\mu}^\dagger(y)$ is the measure on $\mathcal{X}$ with density $x \mapsto \frac{d \kappa}{d c_{(\kappa \circ_m \mu)}}(x, y)$ with respect to $\mu$, in which $c_{(\kappa \circ_m \mu)}$ is the constant kernel with value $\kappa \circ_m \mu$.
It requires the assumption that there exists a dominating measure: for $\mu$-almost all $x \in \mathcal{X}$, $\kappa(x)$ is absolutely continuous with respect to $\kappa \circ_m \mu$.
This is in particular always true in the case of a countable space $\mathcal{X}$, for which the result simplifies to

\begin{lstlisting}
lemma posterior_eq_withDensity_of_countable :
    ∀ᵐ y ∂(κ ∘ₘ μ), (κ†μ) y = μ.withDensity (fun x ↦ (κ x).rnDeriv (κ ∘ₘ μ) y)
\end{lstlisting}
That is, for $\kappa \circ_m \mu$-almost all $y \in \mathcal{Y}$, $\kappa_{\mu}^\dagger(y)$ is the measure on $\mathcal{X}$ defined by $\kappa_{\mu}^\dagger(y)(\{x\}) = \mu(\{x\}) \cdot \frac{d \kappa(x)}{d (\kappa \circ_m \mu)}(y)$ for all $x \in \mathcal{X}$.

\section{Independence}\label{sec:independence}

Sections~\ref{sec:kernels} and \ref{sec:disintegration} described kernels and how they can be disintegrated and used to define conditional distributions.
Those tools are used more widely in \mathlib{} than in most mathematical texts, and in particular they are used to define independence and conditional independence.

\begin{definition}[Independence and conditional independence]\label{def:independence}
Two random variables $X : \Omega \to \mathcal{X}$ and $Y : \Omega \to \mathcal{Y}$ are independent for a measure $P \in \mathcal{M}(\Omega)$ if for all measurable sets $A$ of $\mathcal{X}$ and $B$ of $\mathcal{Y}$, 
\begin{align*}
    P(X^{-1}(A) \cap Y^{-1}(B))
    &= P(X^{-1}(A)) \cdot P(Y^{-1}(B))
    \: .
\end{align*}
Those two random variables are conditionally independent given a sub-$\sigma$-algebra $\mathcal{G}$ of the $\sigma$-algebra of $\Omega$ if for all measurable sets $A$ of $\mathcal{X}$ and $B$ of $\mathcal{Y}$, for $P$-almost every $\omega \in \Omega$, 
\begin{align*}
    P(X^{-1}(A) \cap Y^{-1}(B) | \mathcal{G})(\omega)
    &= P(X^{-1}(A) | \mathcal{G})(\omega) \cdot P(Y^{-1}(B) | \mathcal{G})(\omega)
    \: .
\end{align*}
\end{definition}

There is also a definition of both types of independence for more than two random variables (which is also in \mathlib{}), but we will keep the discussion to two random variables for simplicity.
In the library independence is defined first for sets of sets (intended to be used for $\pi$-systems), then for $\sigma$-algebras by seeing them as sets of sets, and finally for random variables by taking the $\sigma$-algebras they generate.
We will only focus on the case of random variables here, but the same discussion applies to the other two cases.

We introduced a new definition using kernels (\leanlink{https://github.com/leanprover-community/mathlib4/blob/dd465758525e4dcf29027e0ebc5d8fd552be361f/Mathlib/Probability/Independence/Kernel/IndepFun.lean#L55}{Kernel.IndepFun}) and then define both types of independence as particular cases.

\begin{definition}[Kernel independence]\label{def:kernel-independence}
Two random variables $X : \Omega \to \mathcal{X}$ and $Y : \Omega \to \mathcal{Y}$ are conditionally independent given a kernel $\kappa : \mathcal{T} \rightsquigarrow \Omega$ and a measure $\mu \in \mathcal{M}(\mathcal{T})$ if for all measurable sets $A$ of $\mathcal{X}$ and $B$ of $\mathcal{Y}$, we have that for $\mu$-almost every $t \in \mathcal{T}$,
\begin{align*}
    \kappa(t)(X^{-1}(A) \cap Y^{-1}(B))
    &= \kappa(t)(X^{-1}(A)) \cdot \kappa(t)(Y^{-1}(B))
    \: .
\end{align*}
\end{definition}

This definition can recover unconditional independence by taking $\kappa$ to be a constant kernel with value $P$ from a space with one point.
It recovers conditional independence for $\Omega$ a standard Borel space by taking $\mu$ to be the restriction of the measure $P$ to a sub-$\sigma$-algebra and $\kappa$ the conditional expectation kernel $P^{\mathcal{G}}$.
We obtain, almost everywhere,
\begin{align*}
    P^{\mathcal{G}}(\omega)(X^{-1}(A) \cap Y^{-1}(B))
    &= P^{\mathcal{G}}(\omega)(X^{-1}(A)) \cdot P^{\mathcal{G}}(\omega)(Y^{-1}(B))
    \: ,
\end{align*}
and since $P^{\mathcal{G}}(\omega)(X^{-1}(A))$ is almost everywhere equal to $P(X^{-1}(A) | \mathcal{G})(\omega)$ and similarly for the other sets, we recover the definition of conditional independence.

We thus have three definitions of independence in \mathlib{}: independence, conditional independence and independence given a kernel and a measure.
However, the lemmas are all proved only once, for the kernel definition, and only restated with one-line proofs for the two particular cases.

\paragraph{Strengths and weaknesses of the kernel definition}

A drawback to that approach is a slight loss in generality for the conditional independence definition.
Indeed, the existence of the conditional expectation kernel $P^{\mathcal{G}}$ is true only under assumptions on $\Omega$ (we defined it for $\Omega$ a standard Borel space), and thus in \mathlib{} conditional independence is only defined in that case.
However, Definition~\ref{def:independence} does not need any assumption on $\Omega$ and an implementation of conditional independence which does not go through kernels would be more general, and its basic properties would hold in that generality.
Unconditional independence properties could be obtained from the conditional case by taking $\mathcal{G}$ to be the trivial $\sigma$-algebra.

Nonetheless, conditional independence is overwhelmingly used in cases where the conditional expectation kernel exists, often to obtain properties of conditional distributions: for example if $X$ is independent of $Y$ given $Z$, the conditional distribution of $X$ given $(Y,Z)$ is almost surely equal to the conditional distribution of $X$ given $Z$ (see further down).
\citet{forre2021transitional} introduces a generalization of conditional independence. It is defined for any measurable spaces but he makes the choice of requiring the existence of a conditional distribution to declare that two random variables are conditionally independent given another.
Otherwise the property is not deemed useful enough.
It is also notable that manipulating conditional independence is much easier when conditional kernels exists, as explained above at the end of Section~\ref{sub:condDistrib}.
We thus made the choice of accepting the loss of generality for convenience: that choice could be revisited if the need for more general spaces arises.
Another more minor refactor would be to follow the example of \citet{forre2021transitional} and extend our definition to any space by requiring that there exists a conditional kernel such that the definition applies.

The proofs of the properties of independence with respect to a kernel and a measure are thus slightly simplified versions of those of conditional independence.
The abstraction of the conditional expectation into a kernel is also surprisingly helpful in practice.
First, conditional expectations lead to long expressions which can be hard to read in the view of a \Lean{} editor showing the context and goal, and the kernel expression is often more readable.
But besides that minor convenience, the main advantage is that  the rigidity of kernels (due to the different types) prevents us from making mistakes when composing them.
Proofs involving conditional independence would involve several $\sigma$-algebras on the same space, hence a function on a space could be measurable with respect to one $\sigma$-algebra but not another, and it is easy to inadvertently mistake one for the other when attempting to code a proof.
When using kernels, the different types ensure that we cannot compose two kernels unless the types match precisely, and there is only one $\sigma$-algebra involved on each type.
It is often the case in a proof that if two expressions involving kernels have the same type, then they are equal, which makes writing the proofs easier.
That is of course not a mathematical truth but only a practical observation: only deterministic kernels satisfy $(\kappa \parallel_k \kappa) \circ_k \Delta_\mathcal{X} = \Delta_\mathcal{Y} \circ_k \kappa$, although they both have type $\mathcal{X} \rightsquigarrow \mathcal{Y}\times \mathcal{Y}$.

Thus we find that the formalization of some aspects of probability theory is made easier by considering many probability spaces linked by kernels, rather than the standard practice of considering a single probability space on which every random variable is defined.
The relative rigidity of the type theory of \Lean{} can be an obstacle, as the discussion about associativity of composition-products showed, but it can also guide towards the correct statements as it detects badly typed expressions.

Our definition of independence with respect to a kernel and a measure is not an attempt to write the most general possible notion of independence, but merely a minimal abstraction of independence and conditional independence (in standard Borel spaces) that allows us to prove the properties we want only once, with a convenient framework.
Many generalizations exist and we mention only two here, which also focus on kernels.
\citet{forre2021transitional} defines transitional conditional independence, which is independence of two kernels given another kernel, in a transitional probability space (which is defined by a fourth kernel).
We remark that our definition takes inspiration from this and features four objects as well: two random variables, a kernel and a measure.
Both random variables and measures are special cases of kernels, and Forré extended independence to the general case of four kernels.
\citet{cho2019disintegration, fritz2020synthetic} consider categorical definitions of independence of morphisms in Markov categories (and Markov kernels are such a category).

\paragraph{Conditional independence and conditional distributions}

Another characterization of conditional independence in standard Borel spaces is that $X$ and $Y$ are conditionally independent given a third random variable $Z$ (that is, given its generated $\sigma$-algebra) if and only if the conditional distribution of $Y$ given $(X, Z)$ is almost surely equal to the conditional distribution of $Y$ given $Z$.

\begin{lstlisting}
lemma condIndepFun_iff_condDistrib_prod_ae_eq_prodMkRight
    (hX : Measurable X) (hY : Measurable Y) (hZ : Measurable Z) :
    CondIndepFun (mγ.comap Z) hZ.comap_le Y X μ ↔
      condDistrib X (fun ω ↦ (Z ω, Y ω)) μ =ᵐ[μ.map (fun ω ↦ (Z ω, Y ω))]
        (condDistrib X Z μ).prodMkRight _ := by
\end{lstlisting}

The definition \lstinline|CondIndepFun m hm Y X μ| means that $Y$ and $X$ are conditionally independent given the sub-$\sigma$-algebra $m$ of the $\sigma$-algebra of $\Omega$, for the measure $\mu$.
The argument \lstinline|hm| is a proof that $m$ is indeed a sub-$\sigma$-algebra of the $\sigma$-algebra of $\Omega$.
Here the $\sigma$-algebra used is $\sigma(Z)$, the $\sigma$-algebra generated by $Z$, which is obtained by \lstinline|MeasurableSpace.comap Z inferInstance|.
It is a sub-$\sigma$-algebra because $Z$ is measurable, which is proved by \lstinline|hZ.comap_le|.
Writing about conditional independence given a random variable $Z$ instead of a sub-$\sigma$-algebra $\mathcal{G}$ is currently rather verbose, and we could consider adding a more convenient notation for that common case in the future.
The definition \lstinline|Kernel.prodMkRight| turns a kernel $\mathcal{Z} \rightsquigarrow \mathcal{X}$ into a kernel $\mathcal{Z} \times \mathcal{Y} \rightsquigarrow \mathcal{X}$ by ignoring the second argument.

That result was not immediate given the definitions: \lstinline|CondIndepFun| is defined in terms of the conditional expectation kernel $P^{\sigma(Z)}$, which is a \lstinline|condDistrib| of identity functions and that definition needs to be related to conditional distributions of the variables involved.

\section{Sub-Gaussian random variables}\label{sec:subGaussian}

Independence is far from the only property in probability theory that is defined in two versions, conditional and unconditional.
Another example from \mathlib{} is the notion of sub-Gaussian random variables, and we applied the same idea to define them in a unified way.

A sub-Gaussian random variable is a real-valued random variable whose tail decreases at least as fast as a Gaussian.
We refer the reader to \cite{vershynin2018high} for a thorough introduction to sub-Gaussian random variables and their properties.
For $X : \Omega \to \mathbb{R}$ a random variable, the function $u \mapsto P\left[e^{u X}\right]$ is called the moment generating function (MGF) of $X$ (recall that $P[\cdot]$ denotes the expectation under the probability measure $P$).
Formally, $X : \Omega \to \mathbb{R}$ is sub-Gaussian with constant $\sigma^2$ on a probability space $(\Omega, P)$ if $e^{u X}$ is integrable for all $u \in \mathbb{R}$ and the MGF $u \mapsto P\left[e^{u X}\right]$ satisfies the inequality $P\left[e^{u X}\right] \le e^{u^2 \sigma^2/2}$ for all $u \in \mathbb{R}$.
The function $u \mapsto e^{u^2 \sigma^2/2}$ is the MGF of a real Gaussian random variable with mean 0 and variance $\sigma^2$, hence the name sub-Gaussian.
Sub-Gaussianity is a useful property to prove concentration inequalities such as Hoeffding's inequality \cite{hoeffding1963probability}, which states in the independent case that if $X_1, \ldots, X_n$ are independent and sub-Gaussian with constant $\sigma^2$, then for all $t > 0$,
\begin{align*}
    P\left(\sum_{i=1}^n X_i \ge t\right) \le \exp\left(-\frac{t^2}{2n\sigma^2}\right) \: .
\end{align*}
In \mathlib{}, this is \leanlink{https://github.com/leanprover-community/mathlib4/blob/dd465758525e4dcf29027e0ebc5d8fd552be361f/Mathlib/Probability/Moments/SubGaussian.lean#L787}{HasSubgaussianMGF.measure_sum_range_ge_le_of_iIndepFun}.
Concentration inequalities are useful to bound the probability that random variables deviates from their mean, and have many applications in statistics and machine learning.
The Lean definition of sub-Gaussian that we added to \mathlib{} is as follows.

\begin{lstlisting}
structure HasSubgaussianMGF (X : Ω → ℝ) (c : ℝ≥0) (μ : Measure Ω := by volume_tac) : Prop where
  integrable_exp_mul : ∀ t : ℝ, Integrable (fun ω ↦ exp (t * X ω)) μ
  mgf_le : ∀ t : ℝ, mgf X μ t ≤ exp (c * t ^ 2 / 2)
\end{lstlisting}
The \lstinline|:= by volume_tac| code defines a default value in case no measure is provided. That default value uses a tactic (\lstinline|volume_tac|) that infers the measure if $\Omega$ has been endowed with a \lstinline|MeasureSpace| instance. For example if $\Omega = \mathbb{R}$, we do not need to specify that we mean sub-Gaussian with respect to the Lebesgue measure.

The conditional version of the sub-Gaussian property is defined similarly, by replacing the expectation $P[\cdot]$ by the conditional expectation $P[\cdot | \mathcal{G}]$ given a sub-$\sigma$-algebra $\mathcal{G}$ of the $\sigma$-algebra of $\Omega$.
A random variable $X : \Omega \to \mathbb{R}$ is conditionally sub-Gaussian with constant $\sigma^2$ given $\mathcal{G}$ if $e^{u X}$ is integrable for all~$u \in \mathbb{R}$ and $P$-almost surely, $P\left[e^{u X} \mid \mathcal{G} \right] \le e^{u^2 \sigma^2/2}$ for all~$u \in \mathbb{R}$.
We note that this is almost never mentioned as a separate definition in textbooks \cite{boucheron2013concentration, vershynin2018high, lattimore2020bandit}, which implicitly assume that conditional versions of properties are defined by replacing expectations by conditional expectations.
That definition is not however how we defined it in \mathlib{}: we instead introduce a new notion of sub-Gaussian with respect to a kernel and a measure, and define conditionally sub-Gaussian as a particular case of that definition for the conditional expectation kernel.

\begin{definition}\label{def:subGaussian_kernel}
A random variable $X : \Omega \to \mathbb{R}$ is sub-Gaussian with respect to a Markov kernel $\kappa : \mathcal{T} \rightsquigarrow \Omega$ and a measure $\mu \in \mathcal{M}(\mathcal{T})$ if
\begin{itemize}
    \item $\exp (u X)$ is integrable with respect to $\kappa \circ_m \mu$ for all~$u \in \mathbb{R}$,
    \item for $\mu$-almost all $t$, the moment generating function of $X$ under $\kappa(t)$ satisfies, for all~$u \in \mathbb{R}$,
    \begin{align*}
        \kappa(t)\left[e^{u X}\right] \le e^{u^2/2} \: .
    \end{align*}
\end{itemize}
\end{definition}

Again, the definition can also recover the unconditional case by taking $\kappa$ to be a constant kernel from a space with one point.
When we specialize that definition to the conditional expectation kernel $P^{\mathcal{G}}$ and a measure $P$ restricted to $\mathcal{G}$ (denoted by $P_{| \mathcal{G}}$), the two conditions become:
\begin{itemize}
    \item $\exp (u X)$ is integrable with respect to $P^{\mathcal{G}} \circ_m P_{| \mathcal{G}} = P$ for all~$u \in \mathbb{R}$,
    \item for $P_{| \mathcal{G}}$-almost all $\omega \in \Omega$, for all~$u \in \mathbb{R}$,
    $P^{\mathcal{G}}\left[e^{u X(\omega)}\right] \le e^{u^2/2}$~.
\end{itemize}
Since  $P^{\mathcal{G}}\left[e^{u X(\omega)}\right]$ is almost everywhere equal to $P\left[e^{u X} | \mathcal{G}\right](\omega)$ and is a $\mathcal{G}$-measurable function, we recover the definition of conditional sub-Gaussian: for $P$-almost all $\omega \in \Omega$, for all~$u \in \mathbb{R}$, $P\left[e^{u X} \mid \mathcal{G}\right](\omega) \le e^{u^2/2}$~.

An advantage of going through a kernel definition is that this definition is useful in itself, particularly in a context like the one of the Ionescu-Tulcea theorem where the probability space we study comes from a sequence of kernels.

\paragraph{The integrability condition in Definition~\ref{def:subGaussian_kernel}}
In the unconditional case, we require that $\exp(u X)$ is integrable with respect to $P$, which could be interpreted as a technical condition to ensure that the MGF that appears in the inequality is well-defined.
The analogue of that for the kernel case would be to require that for $\mu$-almost all $t \in \mathcal{T}$, for all~$u$, $\exp(u X)$ is integrable with respect to $\kappa(t)$ (and this is how we implemented it originally).
This is implied by the integrability with respect to $\kappa \circ_m \mu$, but is a weaker condition.
Let's call it weak integrability.
Integrability of a measurable function $f$ with respect to $\kappa \circ_m \mu$ is equivalent to $\mu$-almost everywhere integrability of $f$ with respect to $\kappa(t)$ together with integrability of $t \mapsto \int_y \Vert f(y) \Vert d \kappa(t)$.
The weak integrability condition that ensures only that the MGF inequality makes sense has a flaw: it does not compose well.

The issue became apparent when we wanted to prove the following result about sums of sub-Gaussian random variables.
Suppose that we have a measure $\nu$ on $\mathcal{T}$ and two kernels $\kappa : \mathcal{T} \rightsquigarrow \mathcal{X}$ and $\eta : \mathcal{T} \times \mathcal{X} \rightsquigarrow \mathcal{Y}$.
Those measure and kernels may describe the evolution of the state of a system (from a space $\mathcal{T}$ to $\mathcal{X}$ and then $\mathcal{Y}$), which we observe through $\mathbb{R}$-valued random variables $X : \mathcal{X} \to \mathbb{R}$ and $Y : \mathcal{Y} \to \mathbb{R}$.
Then if $X$ is sub-Gaussian with respect to $\kappa$ and $\nu$ and $Y$ is sub-Gaussian with respect to $\eta$ and $\nu \otimes_m \kappa \in \mathcal{M}(\mathcal{U} \times \mathcal{X})$, then $X + Y$ (random variable on the measurable space $\mathcal{X} \times \mathcal{Y}$) is sub-Gaussian with respect to $\kappa \otimes_k \eta : \mathcal{U} \rightsquigarrow \mathcal{X} \times \mathcal{Y}$ and $\nu$.

\begin{lstlisting}
lemma Kernel.HasSubgaussianMGF.add_compProd [IsZeroOrMarkovKernel η]
    (hX : HasSubgaussianMGF X cX κ ν) (hY : HasSubgaussianMGF Y cY η (ν ⊗ₘ κ)) :
    HasSubgaussianMGF (fun p ↦ X p.1 + Y p.2) (cX + cY) (κ ⊗ₖ η) ν
\end{lstlisting}
That lemma is the right tool to prove concentration inequalities for sums of sub-Gaussian random variables with respect to a sequence of kernels.
In order to prove the MGF inequality in that result, we have to obtain at least that for $\nu$-almost all $t \in \mathcal{T}$, for all $u \in \mathbb{R}$, $\exp(u (X + Y))$ is integrable with respect to $(\kappa \otimes_k \eta)(t) = \kappa(t) \otimes_m \eta$, in order for the MGF to be defined.
With the weak integrability condition, we only know that for $\nu$-almost all $t \in \mathcal{T}$, for all $u$, $\exp(u X)$ is integrable with respect to $\kappa(t)$ and that for $(\nu \otimes_m \kappa)$-almost all $(t, x) \in \mathcal{T} \times \mathcal{X}$, for all $u$, $\exp(u Y)$ is integrable with respect to $\eta(t, x)$.
That is not enough to conclude, mainly because the result of integrability with respect to $\kappa(t) \otimes_m \eta$ is stronger than $\kappa(t)$-almost everywhere integrability with respect to $\eta(t,x)$.
The stronger integrability condition in Definition~\ref{def:subGaussian_kernel} solves that issue.

\paragraph{General principle}

The general principle we applied both in the independence and sub-Gaussian cases is simple: where the non-conditional case features an expectation $\mu[\cdot]$ and the conditional case an almost sure statement about a conditional expectation $\mu[\cdot | \mathcal{G}]$, we introduce a kernel $\kappa : \mathcal{T} \rightsquigarrow \Omega$ and a measure $\nu \in \mathcal{M}(\mathcal{T})$ and replace the expectation by $\kappa(t)[\cdot]$ for $\nu$-almost all $t \in \mathcal{T}$.
The unconditional case is recovered by taking $\mathcal{T} = \mathbf{1}$ (space with one point) and $\kappa$ the constant kernel from that space to $\Omega$.
The conditional case is recovered in Borel spaces by taking $\mathcal{T} = \Omega$, $\nu$ the restriction of $\mu$ to $\mathcal{G}$ and $\kappa = P^{\mathcal{G}}$ the conditional expectation kernel.

A property of the type ``for $\mu$-almost all $x$, $P(x)$'' would be generalized to ``for $\nu$-almost all $t \in \mathcal{T}$, for $\kappa(t)$-almost all $x \in \mathcal{X}$, $P(x)$'', which is equivalent to ``for $\nu \otimes_m \kappa$-almost all $(t, x) \in \mathcal{T} \times \mathcal{X}$, $P(x)$''.
The latter formulation is more convenient to use because it features a single ``almost all'' instead of two nested ones.

\section{Information theory and divergences}\label{sec:info}

In this last section on applications of kernels to the formalization of other probabilistic notions, we will discuss information-theoretic quantities: entropy and the Kullback-Leibler divergence (KL).
Those examples are not from \mathlib{}, but from contributions of the author to other projects (although the definition of KL is now in \mathlib{}, \leanlink{https://github.com/leanprover-community/mathlib4/blob/dd465758525e4dcf29027e0ebc5d8fd552be361f/Mathlib/InformationTheory/KullbackLeibler/Basic.lean#L57}{klDiv}).
They should be seen as experiments on how the kernel approach can be applied to other notions, and not as definitive design choices.

In the PFR project \cite{Anderson_Formalization_of_the_2023}, we formalized the entropy of a probability measure $\mu$ on a finite set $\mathcal{X}$ as $H(\mu) = - \sum_{x \in \mathcal{X}} \mu(\{x\}) \log \mu(\{x\})$, and the conditional entropy of a kernel $\kappa : \mathcal{X} \rightsquigarrow \mathcal{Y}$ with respect to a probability measure $\mu$ on $\mathcal{X}$ as
\begin{align*}
H_k(\kappa \mid \mu) = \sum_x \mu(x) H(\kappa(x)) \: .
\end{align*}
Conditional entropy of random variables in a probability space $(\Omega, P)$ was given a direct definition
\begin{align*}
H(X|Y) = -\sum_y \sum_x P(Y=y) P(X=x \mid Y=y) \log(P(X=x|Y=y))
\: ,    
\end{align*}
but we also proved that the conditional entropy of $X$ given $Y$ is equal to $H_k\left( P_{X|Y} \mid P_Y \right)$ (recall that $P_{X|Y}$ is a conditional distribution, see Section~\ref{sub:condDistrib}) and derived many properties for the kernel version of entropy, before then specializing to the conditional entropy.
For example, the chain rule for kernel entropy can be stated as
\begin{align*}
    H_k(\kappa \otimes_k \eta \mid \mu) = H_k(\kappa \mid \mu) + H_k(\eta \mid \mu \otimes_m \kappa)
    \: .
\end{align*}
The advantage of using the kernel approach there is mostly in the clarity of the proofs: abstracting away the conditional distribution into a generic kernel streamlines the reasoning.

The Kullback-Leibler divergence definition now in \mathlib{} was developed in the TestingLowerBounds project \cite{TestingLowerBounds}.
A highlight of the project is a proof of the data-processing inequality for f-divergences, a class of divergences between probability distributions that includes KL, as well as the chain rule for KL.
The Kullback-Leibler divergence between two probability measures $\mu, \nu$ on a measurable space $\mathcal{X}$ is defined as
\begin{align*}
    \KL(\mu, \nu) = \int_{x \in \mathcal{X}} \log\left(\frac{d \mu}{d \nu}(x)\right) d \mu
    \: ,
\end{align*}
if $\mu$ is absolutely continuous with respect to $\nu$, and $+\infty$ otherwise.
The usual formulation of the chain rule for $\KL$ states that if $\mu, \nu$ are two probability measures on $\mathcal{X}$ and $\kappa, \eta : \mathcal{X} \rightsquigarrow \mathcal{Y}$ are two Markov kernels, then
\begin{align*}
    \KL(\mu \otimes_m \kappa, \nu \otimes_m \eta)
    &= \KL(\mu, \nu) + \KL(\kappa, \eta \mid \mu)
    \: ,
\end{align*}
in which $\KL(\kappa, \eta \mid \mu) = \int_{x \in \mathcal{X}} \KL(\kappa(x), \eta(x)) d \mu$ is the conditional KL divergence.
The chain rule proof then required the function $x \mapsto \KL(\kappa(x), \eta(x))$ to be measurable, which is true in spaces in which a kernel Radon-Nikodym derivative of $\kappa$ with respect to $\eta$ exists.
A version of that same result that avoids the measurability issue and does not require any assumption on the measurable spaces is
\begin{align*}
    \KL(\mu \otimes_m \kappa, \nu \otimes_m \eta)
    &= \KL(\mu, \nu) + \KL(\mu \otimes_m \kappa, \mu \otimes_m \eta)
    \: .
\end{align*}
Expressing the conditional divergence with composition-products of kernels instead of an integral avoids assumptions and makes manipulations easier.
Similarly, the fact that ``conditioning increases divergence'' usually refers to the inequality $\KL(\kappa \circ_m \mu, \eta \circ_m \mu) \le \KL(\kappa, \eta \mid \mu)$, but the version $\KL(\kappa \circ_m \mu, \eta \circ_m \mu) \le \KL(\mu \otimes_m \kappa, \mu \otimes_m \eta)$ is more general.
In fact, any divergence (a function $D$ that maps two probability measures $\mu, \nu$ to a value $D(\mu, \nu) \in \mathbb{R}_{+, \infty}$ with $D(\mu, \mu) = 0$) that satisfies the important data-processing inequality $D(\kappa \circ_m \mu, \kappa \circ_m \nu) \le D(\mu, \nu)$ also satisfies the ``conditioning increases divergence'' inequality for the composition-product of kernels version.
In that generality, the composition-product version may not be equal to the integral version, although equality holds for the large class of f-divergences.

For the Rényi divergence $R_\alpha(\mu, \nu) = \frac{1}{\alpha - 1} \log \left( \int_x \left(\frac{d\mu}{d(\mu+\nu)}(x)\right)^\alpha \left(\frac{d\nu}{d(\mu+\nu)}(x)\right)^{1 - \alpha} d(\mu + \nu) \right)$ for $\alpha \in (0,1)$, the composition-product and integral definitions do not coincide.
It is remarkable that the conditional definition that enjoys a form of the chain rule is not the integral definition, but the composition-product of kernels version $R_\alpha(\mu \otimes_m \kappa, \nu \otimes_m \eta)$ \cite[Chapter 7]{polyanskiy2025information}.
We thus expect Markov kernels to play an important role in the formalization of probability divergences in \mathlib{}.

\section{Related work}

We discuss formalizations of transition kernels in other proof assistants, as well as an application of kernels in \mathlib{} to prove the Ionescu-Tulcea theorem and define the product measure of an infinite family of probability measures.

\subsection{Markov kernels in other proof assistants}

The formalization of Markov kernels in other proof assistants is mostly motivated by the semantics of probabilistic programming languages \cite{staton2020probabilistic}, while our goal is to provide a general-purpose theory of kernels for probability theory.
In Isabelle/HOL, \citet{hirata2023semantic} formalized quasi-Borel spaces and defined Markov kernels, finite and s-finite kernels.
In Rocq/Coq, \citet{affeldt2025semantics} also formalized Markov, finite and s-finite kernels, and additionally $\sigma$-finite kernels.
They discuss in details the problem of proving the measurability needed to define the composition of kernels, which we omitted in this paper.
Both of those formalizations build classes of kernels and some of their compositions (corresponding to what we present in Section~\ref{sec:kernels}) but do not discuss disintegration of kernels or conditional distributions.

A small difference between the implementations of both \citet{hirata2023semantic} and \citet{affeldt2025semantics} and ours is that they do not define a kernel as a measurable function into a measurable space of measures, but as a function into measures that satisfies the set-based measurability condition directly (for every measurable set $B$, the function $x \mapsto \kappa(x)(B)$ is measurable).
Our definition uses the measurable space on measures to write that the function $x \mapsto \kappa(x)$ is measurable instead of spelling out what it means in terms of evaluation on measurable sets.

In our implementation, only \lstinline|Kernel| is a structure, and then s-finite, finite and Markov kernels are properties of \lstinline|Kernel| (which are implemented as typeclasses).
In contrast, \citet{affeldt2025semantics} define structures for each class of kernels they introduce. We are not knowledgeable enough about the Rocq language to comment meaningfully on the significance of that design choice.

\citet{affeldt2025semantics} mention several composition operations on kernels. Our composition-product is the product of their Theorem~8.1, while our composition corresponds to the operation of their Theorem~8.3.
Interestingly, the central operation in their work (Theorem~5.1) is not exactly one we implemented. It takes two s-finite kernels $\kappa : \mathcal{X} \rightsquigarrow \mathcal{Y}$ and $\eta : \mathcal{X} \times \mathcal{Y} \rightsquigarrow \mathcal{Z}$ and produces a kernel $\kappa ; \eta : \mathcal{X} \rightsquigarrow \mathcal{Z}$, which is such that for all measurable functions $f : \mathcal{Z} \to \mathbb{R}_+$, for all $x \in \mathcal{X}$,
\begin{align*}
    \int_z f(z) d(\kappa ; \eta)(x)
    = \int_{y \in \mathcal{Y}} \int_{z \in \mathcal{Z}} f(z) d \eta(x, y) d \kappa(x) \: .
\end{align*}
This is also the main composition operation discussed by \citet{staton2020probabilistic}. We could implement it by taking the projection of the composition-product to the second coordinate ($(\kappa \otimes_k \eta)_{\mathrm{snd}}$), but we did not see a need for it in our work. This is probably due to our focus on kernels as a tool for probability theory, while those works are motivated by probabilistic programming languages.

\subsection{The Ionescu-Tulcea theorem}

We highlight a recent application of the kernel framework: the Ionescu-Tulcea theorem, which was implemented in \Lean{} and added to \mathlib{} by \citet{marion2025formalization}.

Markov kernels arise naturally in the description of the sequential evolution of a system.
For example, in stochastic optimization, an algorithm sequentially queries a function $f: \mathbb{R}^d \to \mathbb{R}$ at points $X_1, X_2, \ldots$ that are chosen based on the results of previous queries, with the goal of finding a minimizer of the function.
A query returns a noisy observation $Y_i$ of the function value $f(X_i)$ (or possibly additional information like its gradient).

The conditional distribution of $Y_i$ given $X_i$ is naturally described by a Markov kernel $\kappa_i : \mathbb{R}^d \rightsquigarrow \mathbb{R}$.
The choice of $X_{i+1}$ based on the previous queries is also described by a Markov kernel $\eta_i : (\mathbb{R}^d \times \mathbb{R})^i \rightsquigarrow \mathbb{R}^d$, and we thus naturally end up with a chain of kernels to describe the interaction.
We will want to make global probabilistic statements about the optimization procedure. For example, we may want to prove that with probability 1, the sequence of queries converges to a minimizer of $f$.
To do that, we need to define a probability space on which all the random variables $X_i$ and $Y_i$ are defined.
The Ionescu-Tulcea theorem gives such a probability space.
We can see it as a tool to go from the kernel point of view in which we describe transitions between multiple measurable spaces to the single probability space point of view, in which every random variable is defined on the same space.

\begin{theorem}[Ionescu-Tulcea]\label{thm:ionescu-tulcea}
Let $(X_n)_{n \in \mathbb{N}}$ be a family of measurable spaces. Let $(\kappa_n)_{n \in \mathbb{N}}$ be a family of Markov kernels such that for any $n$, $\kappa_n$ is a kernel from $\prod_{i=0}^n X_{i}$ to $X_{n+1}$.
Let $\pi_{[1,n]} : \prod_{i=1}^{\infty} X_i \to \prod_{i=1}^n X_i$ be the projection on the first $n$ coordinates.
Then there exists a unique Markov kernel $\xi : X_0 \rightsquigarrow \prod_{i = 1}^{\infty} X_{i}$ such that for any $n \ge 1$, the pushforward of $\xi$ by $\pi_{[1,n]}$ is
$\pi_{[1,n]*} \xi = \kappa_0 \otimes_k \ldots \otimes_k \kappa_{n-1}$.
\end{theorem}

This theorem applies to the stochastic optimization setting described above with kernels $\eta_i \otimes_k \kappa_i : (\mathbb{R}^d \times \mathbb{R})^i \rightsquigarrow \mathbb{R}^d \times \mathbb{R}$ (where we abuse notation and see $\kappa_i$ as a kernel from $(\mathbb{R}^d \times \mathbb{R})^i \times \mathbb{R}^d$ to $\mathbb{R}$ that ignores the first argument).

In the formalization of that Theorem, an important difficulty is the lack of associativity of products of types, as explained in Section~\ref{sub:composition} about the composition-product.
The article \cite{marion2025formalization} discusses this issue in detail and presents a formalization strategy to avoid inserting associators everywhere.

The main application of the Ionescu-Tulcea theorem in \mathlib{} is to define the product measure of an infinite family of probability measures (\leanlink{https://github.com/leanprover-community/mathlib4/blob/dd465758525e4dcf29027e0ebc5d8fd552be361f/Mathlib/Probability/ProductMeasure.lean#L349-L359}{Measure.infinitePi}), indexed by an arbitrary, possibly uncountable, type.

\section{Conclusion}

We presented several applications of Markov kernels to the formalization of probability theory in \mathlib{} and in other projects.
We first described the definition of kernels and their operations, and how they can be used to define conditional distributions through disintegration.
We then showed how we introduced a kernel-centric definition for independence and conditional independence, and how a similar approach was used to work with sub-Gaussian random variables.
We expect that approach to see more use in \mathlib{} in the future.

An important area that needs to see more development is the link between the probability space defined by the Ionescu-Tulcea theorem and the structure of the sequence of kernels that generated it.
For example, we can describe a Markov chain by a sequence of kernels that depend only on the last state (and not on the whole history), and we can feed that sequence to the Ionescu-Tulcea theorem to obtain a probability space.
In that space, the Markov chain property will be described by a list of conditional independence statements between random variables.
But the lemmas required to turn a structure on the sequence of kernels into conditional independence properties in the probability space are not yet implemented.
This needs to be done in order to model interactions like the stochastic optimization example we described in Section~\ref{sec:kernels}, or to formalize sequential algorithms like the ones used in machine learning.

\section*{Acknowledgements}
The author would like to thank Sébastien Gouëzel, Etienne Marion, and Kexing Ying for their valuable code reviews and suggestions, which prompted several improvements over the initial implementation of kernels and their applications.
We would also like to acknowledge the collaborative nature of the construction of \mathlib{}: this work is only possible because it builds on the prior efforts of many contributors over several years to build a comprehensive library for mathematics in \Lean{}, notably in measure theory.

\printbibliography

\end{document}

%% file: preamble.tex
\usepackage{newunicodechar}
\usepackage{xspace}
\usepackage{colonequals}
\usepackage[colorlinks=true]{hyperref}

\newcommand*{\mathlib}{\textsc{Mathlib}\xspace}
\newcommand*{\Lean}{\textsc{Lean}\xspace}

\newcommand{\KL}{\mathrm{KL}}
\newcommand{\id}{\mathrm{id}}

\newunicodechar{≪}{\ensuremath{\ll}}
\newunicodechar{ᵐ}{\ensuremath{^m}}
\newunicodechar{ₘ}{\ensuremath{_m}}
\newunicodechar{ˢ}{\ensuremath{^s}}
\newunicodechar{ₖ}{\ensuremath{_k}}
\newunicodechar{𝓧}{\ensuremath{\mathcal{X}}}
\newunicodechar{𝓨}{\ensuremath{\mathcal{Y}}}
\newunicodechar{𝓩}{\ensuremath{\mathcal{Z}}}